\documentclass[a4paper, reprint, showkeys, nofootinbib, twoside,superscriptaddress]{revtex4-1}
\usepackage{amsmath,amssymb,amsfonts}
\usepackage[english]{babel}
\usepackage[utf8]{inputenc}

\usepackage[table]{xcolor}
\usepackage{amsthm}
\usepackage{mathtools}
\usepackage{graphicx}
\usepackage[left=23mm,right=13mm,top=35mm,columnsep=15pt]{geometry} 
\usepackage{adjustbox}
\usepackage{placeins}
\usepackage[T1]{fontenc}
\usepackage{lipsum}
\usepackage{csquotes}

\usepackage{float}
\usepackage{url}

\usepackage{pifont}

\usepackage{bm}
\usepackage{booktabs}
\usepackage{hyperref}
\newcommand{\dd}{\mathrm{d}}

\allowdisplaybreaks
\begin{document}

% keep the below 6 commands only to have numbering in actual paper match the appendix numbering here 
%\renewcommand{\thesection}{\arabic{section}}
%\renewcommand{\thesubsection}{\arabic{subsection}}
%\renewcommand{\thesubsubsection}{\arabic{subsubsection}}

%\makeatletter
%\renewcommand{\p@subsection}{\thesection.}
%\renewcommand{\p@subsubsection}{\thesection.\thesubsection.}
\makeatother

\title{Revisiting the Excess Volatility Puzzle \\ Through the Lens of the Chiarella Model}

\author{Jutta G. Kurth}
\affiliation{CFM Chair of Econophysics and Complex Systems, \'Ecole polytechnique, 91128 Palaiseau Cedex, France}
\affiliation{LadHyX UMR CNRS 7646, \'Ecole polytechnique, 91128 Palaiseau Cedex, France}

\author{Adam A. Majewski}
%\email{------------}
%\affiliation{LadHyX UMR CNRS 7646, \'Ecole polytechnique, 91128 Palaiseau Cedex, France}
\affiliation{Capital Fund Management, 23 Rue de l'Universit\'e, 75007 Paris, France}
\author{Jean-Philippe Bouchaud}
\affiliation{CFM Chair of Econophysics and Complex Systems, \'Ecole polytechnique, 91128 Palaiseau Cedex, France}
\affiliation{Capital Fund Management, 23 Rue de l'Universit\'e, 75007 Paris, France}
\affiliation{Acad\'emie des Sciences, 23 Quai de Conti, 75006 Paris, France\smallskip}

\date{May 9, 2025} % Leave empty to omit a date

\begin{abstract}
We amend and extend the Chiarella model of financial markets to deal with arbitrary long-term value drifts in a consistent way. This allows us to improve upon existing calibration schemes, opening the possibility of calibrating {\it individual} monthly time series instead of classes of time series. The technique is employed on spot prices of four asset classes from ca. 1800 onward (stock indices, bonds, commodities, currencies). The so-called fundamental value is a direct output of the calibration, which allows us to (a) quantify the amount of excess volatility in these markets, which we find to be large (e.g. a factor $\approx 4$ for stock indices) and consistent with previous estimates; and (b) determine the distribution of mispricings (i.e. the difference between market price and value), which we find in many cases to be bimodal. Both findings are strongly at odds with the Efficient Market Hypothesis. We also study in detail the ``sloppiness'' of the calibration, that is, the directions in parameter space that are weakly constrained by data. The main conclusions of our study are remarkably consistent across different asset classes, and reinforce the hypothesis that the medium-term fate of financial markets is determined by a tug-of-war between trend followers and fundamentalists. 
\end{abstract}


\maketitle
%\tableofcontents
% \input{sections/0intro.tex} 

% \textbf{Context+Stylised facts}
\section{Introduction}\label{sec:intro}
%\addcontentsline{toc}{section}{\nameref{sec:intro}}

The Efficient Market Hypothesis maintains that market prices closely follow fundamental values at all times. Although still a cornerstone of Financial Economics and fiercely defended by some scholars,\footnote{see, e.g., this \href{https://conversableeconomist.com/2025/01/08/interview-with-eugene-fama-for-whom-are-financial-markets-efficient/}{interview} for E. Fama's latest quips on this topic. Chiarella's model can be seen as an explicit alternative to EMH, contradicting Fama's peeve: {\it Now the problem is that behavioral finance doesn’t have any models of its own. It’s just a criticism of other models.}} contradicting evidence has accumulated since the early eighties. Among the most inconvenient facts are (i) Shiller's excess volatility puzzle \cite{shiller1981stock}, i.e. that market volatility appears to be much too high to be explained by the volatility of fundamental values and (ii) the well-documented trend following anomaly, that is, the statistically significant, persistent and profitable positive correlation between past trends and future trends, across all asset classes \cite{asness2013value,lemperiere2014two, hurst2017century} and refs. therein. Such correlations should not exist if markets were efficient. 

A competing theory, that has gained momentum (no pun intended) over the past decades, is the order-driven view of market prices \cite{bouchaud2018trades, gabaix2021search, bouchaud2022inelastic, van2024ponzi}, a.k.a. the inelastic market hypothesis. In such a picture, prices are ``mechanically'' impacted by order flow, independently of fundamental value. Excess buy (resp. sell) pressure, even uninformed, makes prices go up (resp. down), and such an impact persists over the medium to long term. Therefore, accounting for price movements mostly means understanding order flows. Of course, the reasons why people buy or sell are multifarious and based on an infinite variety of incentives and trading signals. In order to model such a complex ecology of market participants \cite{farmer2002market}, Carl Chiarella, and several authors after him, have proposed to retain only three main categories of traders \cite{chiarella1992dynamics, lux1999scaling, schmitt2017bimodality,  majewski2020co, scholl2021marketecology}: ``trend followers'' (who buy/sell when the price has gone up/down), ``fundamentalists'' (who sell/buy when the price is above/below their perceived fair value), and ``noise traders'' (who buy or sell for random reasons, i.e. all other reasons not captured by trend or value).  Chiarella's model and its generalisations offer the simplest Heterogenous Agent Based framework that captures important stylized facts of financial markets, including the excess volatility puzzle, volatility clustering, Black's ``factor 2'' of persistent mispricings \cite{black1986noise,bouchaud2017black}, as well as the long-term ecological coexistence of such strategies. 

Using Bayesian filtering techniques, a full-fledged calibration of the Chiarella model was undertaken for a variety of assets in \cite{majewski2020co}, and clearly supported the co-existence of trend-following and value mean-reversion in most markets, including the possibility to a bimodal distribution of mispricings -- meaning that markets have a higher probability of being over- or under-valued than correctly valued. However, our attempt to extend such a calibration scheme to single stocks revealed problems and inconsistencies. The aim of the present study is to provide a new, more consistent specification of the Chiarella model, and its calibration of the same universe of assets as in \cite{majewski2020co}. This allows us to discuss in more detail the excess volatility puzzle and the issue of bimodality. We also apply the ``sloppiness analysis'' proposed by Sethna et al. \cite{gutenkunst2007sloppy} to our calibration procedure, allowing us to identify the most important features of the model that the data is able to identify. The extension of our method to single stocks, as well as possible further generalisations of the Chiarella model, will be detailed in forthcoming publications. 

The outline of this paper is as follows: in Sec.~\ref{sec:model} the model is introduced and its possible dynamical phases are analytically derived. Sec.~\ref{sec:data} describes the time series data used in this study on which the model is calibrated, detailed in Sec.~\ref{sec:calibration}. The calibration results allow an investigation of excess volatility. To what extent prices typically depart from values is detailed in Sec.~\ref{sec:mispricing_dist} through the mispricing distribution, in particular its bimodality. The ``sloppy'' character of our calibration is discussed in Sec.~\ref{sec:sloppy}. Finally, a conclusion with an outlook is provided in Sec.~\ref{sec:conclusion}. More technical material and supplementary empirical analyses are provided in Appendices.

\section{A Modified Chiarella Model}
\label{sec:model}

\subsection{Model Specification}

We assume the evolution of (log-)prices $P_t$ is governed by linear price impact as suggested in Kyle's seminal work \cite{kyle1985}. Although a linear price impact is ruled out on daily or intraday time scales (see, e.g \cite{bouchaud2018trades}), it is thought to be appropriate on longer, monthly time scales \cite{gabaix2021search,bouchaud2022inelastic} which are of interest in the present work. This means that a price change in a (long enough) time interval $[t, t + \Delta t)$ is proportional to the total signed volume traded in that interval, where the total signed volume is represented by a cumulative demand imbalance $D(t, t+\Delta t)$, i.e.
\begin{equation}
    P_{t+\Delta t} - P_t = \lambda D(t, t +\Delta t).
\end{equation}
Here $\lambda$ is Kyle's lambda, which is inversely proportional to the liquidity of the traded asset. Thus, a product is considered liquid if its price change resulting from a certain traded volume or demand imbalance is relatively small.     

The aggregate demand of all investors is of course diverse and abundant. However, studies reveal that the two types of market participants accounting for a large share of demand imbalances are (a) fundamental value investors and (b) chartists or trend followers (TFs), as done in \cite{chiarella1992dynamics, lux1998, lux1999scaling} and empirically confirmed in \cite{boswijk2007}.

For this reason, a model inspired by \cite{chiarella1992dynamics} was proposed in \cite{majewski2020co}, but whose analytical shortcomings we seek to alleviate in this new study. Like in the latter paper, the HABM-type model studied here contains three groups of investors or \textit{agents}:\footnote{For other HABM specifications, and their calibration on the S\&P500, see the mini-review of T. Lux \cite{lux2021can}.}
\begin{enumerate}
    \item \textbf{Fundamentalists}: investors who believe in a rational fundamental (log-)value $V_t$ of a financial asset. They tend to only step-in when the (log-)price $P_t$ is far away from its value $V_t$: they buy when assets are underpriced ($P_t<V_t$) and sell when they are overpriced ($P_t>V_t$). Fundamentalists' cumulative demand is proportional to the mispricing or price distortion $\delta_t:=V_t-P_t$ with a factor $\Tilde{\kappa}$ quantifying their weight in the market:
    $$D^\textup{F} (t, t+\Delta t) = \Tilde{\kappa} \int_t^{t+\Delta t} (V_s -P_s) \dd s.$$
    We will use below the quantity $\kappa:= \lambda \Tilde{\kappa}$.

    Usually, value traders resort to fundamental analysis for their valuation and we will model value dynamics with an arbitrary (non-stationary) drift, $g_t$, plus random changes $\sigma_V \dd W_t^V$:
    \begin{equation}
    \dd V_t = g_t \dd t + \sigma_V \dd W_t^V \nonumber.
    \end{equation}
    The drift $g_t$ describes the long-term evolution of fundamental value and one should expect its variations to be slow; any faster changes in value being captured by the random term $\dd W_t^V$.
    
    \item \textbf{Trend Followers}: TFs' trading behaviour is independent of a notion fundamental value. Instead, their investment choices rely solely on past price dynamics. They buy, if the price moved up (relative to its long time drift) in the recent past, such that the trend signal $M_t>0$ and sell if the price went down, $M_t<0$. Popular choices for such trend signals are exponentially weighted moving averages (EWMAs) of past returns. The demand is given by an increasing function of the signal $M$ that saturates for strong signals $|M|$ (due to e.g. budget constraints or risk aversion \cite{lemperiere2014two}). With $\Tilde{\beta}$ the weight of TFs and $\gamma$ the signal saturation sensitivity, we posit that the demand of TFs reads
    $$D^\textup{TF}(t, t+\Delta t) = \Tilde{\beta} \int_t^{t+\Delta t} \tanh (\gamma M_s) \dd s.$$ 
    We will use below the quantity $\beta:= \lambda \Tilde{\beta}$.
    
    \item \textbf{Noise Traders}: NTs subsume all those traders that follow strategies uncorrelated to the previous two. One may think of retail investors, or investors who trade on other signals or time horizons. Their cumulative demand is modelled as a Brownian motion $ \sigma 
    _\textup{N}W_t^\textup{N}$, where the standard deviation $\sigma_\textup{N}$ describes their impact in the asset, i.e.
    $$D^\textup{NT}(t, t+\Delta t) = \Tilde{\sigma}_N \int_t^{t+\Delta t} \dd W_s^\textup{N}, $$
    where $\lambda\Tilde{\sigma}_N =: \sigma_N$. The standard Brownian motion $W_t^N$ is independent from random, fast changes of fundamental value described by $W_t^V$.
\end{enumerate}
Thus, the overall cumulative demand leading to a price change is\footnote{One may question whether the demand of Fundamentalists and of Trend Followers is given by their signal or by the change thereof. We restrict here to the original Chiarella specification and will discuss this question further in a subsequent publication.}
\begin{equation}
    D(t, t+\Delta t) = \sum_{i \in I}D^\textup{i} (t, t+\Delta t), 
    \label{eq: demand_imb}
\end{equation}
where $I$ is the set of investor types ($I = \{ F, TF, NT\}$).

In the following we assume that the trend signal $M_t$ is computed as an exponential moving average with a forget rate $\alpha$. Consequently, for $\Delta t \to 0$, the price dynamics is described by the following set of stochastic differential equations
\begin{align}
\label{eq: ModifiedChiarella}
    \dd P_t &= \kappa (V_t-P_t) \dd t + \beta \tanh (\gamma M_t) \dd t + g_t \dd t + \sigma_N \dd W_t^N\nonumber \\
    \dd M_t &= -\alpha M_t \dd t+ \alpha (\dd P_t -g_t  \dd t) \\
    \dd V_t &= g_t \dd t + \sigma_V \dd W_t^V.\nonumber
\end{align}
The parameters $\alpha$, $\kappa$, $\beta$, $\gamma$ are all non-negative.

The quantity $M_t$ denotes the trend signal, which is an EWMA of past {\it drift-adjusted} log-returns. In other words: we assume that trend followers react only to the excess returns of the asset, and do not trend on the secular drift. This would lead to absurd instabilities in the model we want to avoid. As such, this specification is an improvement over the model proposed in \cite{majewski2020co}, where a \textit{constant} drift $g$ appears only in the fundamental value $V$ and not in the definition of $M_t$, nor in the dynamics of $P_t$. As will be shown next, the stability and dynamical phases of the system now become independent of the drift $g_t$, whereas in Ref. \cite{majewski2020co} the dynamical analysis was only valid for $g=0$. This renders the dynamics and calibration improper for assets whose value can strongly drift upwards or downwards, like stocks. In these cases, the specification of Ref. \cite{majewski2020co} may lead to a divergence between $P_t$ and $V_t$ entirely caused by $g_t$ and not as a consequence of demand imbalances. 

A way to mitigate such an effect and prevent the long-term divergence between price and value is to consider, as was done in Ref. \cite{majewski2020co}, a non-linear anchoring term of the form $\kappa_3 (V_t - P_t)^3$. The calibration of such a non-linear model however comes at a higher computational cost and it is important to have a linear model that makes sense and can be meaninfully calibrated before considering a non-linear extension (which we will have to do anyway for reasons explained below).

\subsection{Non-linear Model}\label{sec:NonlinearModel}

Schmitt and Westerhoff \cite{schmitt2017bimodality} and Majewski et al. \cite{majewski2020co} further introduced a model with a non-linear demand function for fundamentalists. They argue and show that the linear fundamentalists' demand is not able to capture the complex nature of value investing, most importantly the uncertainty of investors about fundamental value, which cannot be directly observed but only estimated. It seems reasonable that the reaction of investors is not proportional to mispricing $\delta = V-P$, but much weaker for small mispricings (in view of the uncertainty, leading to an almost flat curve for small $\delta$) and much stronger when mispricing becomes conspicuous \cite{majewski2020co}.

In order to accommodate these departures from linearity, a cubic demand term can be added to the linear term. Within our adjustments the model then reads
\begin{align}
\label{eq: ModifiedChiarellaNonlinear}
    \dd P_t &= f (V_t-P_t) \dd t + \beta \tanh (\gamma M_t) \dd t + g_t \dd t + \sigma_N \dd W_t^N\nonumber \\
    \dd M_t &= -\alpha M_t \dd t+ \alpha (\dd P_t -g_t  \dd t) \\
    \dd V_t &= g_t \dd t + \sigma_V \dd W_t^V,\nonumber
\end{align}
where $f(x) = \kappa x + \kappa_3 x^3$ describes the modified demand. For the model to be compatible with strong mean reversion for large mipricing, one must impose $\kappa_3 >0$.

\subsection{Linear Stability and Bifurcation Analysis} \label{sec:stability_analysis}

The deterministic equivalent to system \eqref{eq: ModifiedChiarella}, which can be analytically studied using methods from dynamical systems theory\footnote{See, e.g., Strogatz \cite{strogatz2018nonlinear} for a comprehensive introduction to non-linear dynamics, Guckenheimer \cite{guckenheimer2013nonlinear} for a comprehensive study of dynamical systems or Lorenz \cite{lorenz1993nonlinear} for theory accompanied by applications in the economic context.} is obtained by letting $\sigma_N=\sigma_V =0$.
Since the mispricing $\delta = P-V$ associated with the system \eqref{eq: ModifiedChiarella} is independent of the drift $g_t$, it is mathematically convenient to study price \textit{relative to} value in $\delta$-$M$-space, effectively reducing the dimensionality by one. The linear system ($\kappa_3 = 0$) then reads
\begin{align}
\label{eq:chiarella_on_mispricing}
    \dot{\delta}_t &= -\kappa\delta_t + \beta \tanh (\gamma M_t) \notag \\
    \dot{M}_t &= -\alpha M_t + \alpha \dot{\delta} \\
    &= -\alpha M_t + \alpha (-\kappa\delta + \beta \tanh (\gamma M_t)).  \notag
\end{align}
The $\delta$-nullcline is
\begin{equation}
    \delta_t = \frac{\beta}{\kappa} \tanh (\gamma M_t)
\end{equation}
and the $M$-nullcline is
\begin{equation}
    \delta_t = \frac{\beta}{\kappa} \tanh (\gamma M_t) -\frac{M_t}{\kappa},
\end{equation}
which is a sigmoidal function for small $|M_t|$ thanks to the $\tanh$ but for large $\pm M_t$ the function diverges to $\mp \infty$.

\begin{figure*}[ht]
    \centering
    \includegraphics[width=0.7\linewidth]{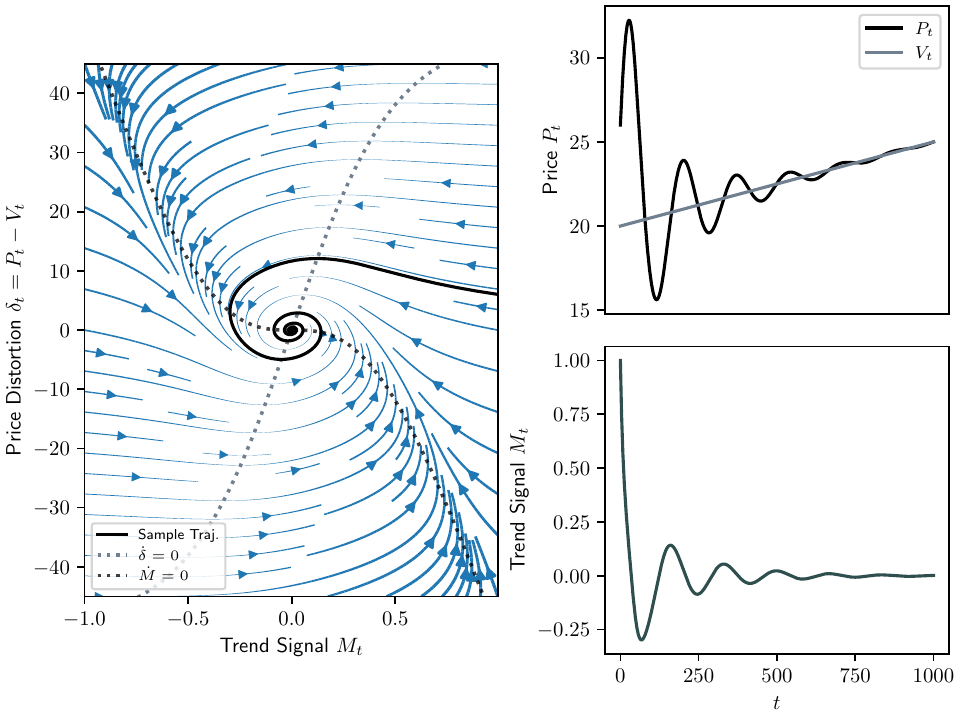}
    \caption{Typical dynamics of system~\eqref{eq: ModifiedChiarella} in the case where its limit set is a spiral, $\kappa > \alpha (\beta\gamma -1)$, and without noise ($\sigma_N=\sigma_V=0$). The parameters are $(\kappa, \, \alpha, \, \beta, \, \gamma) = (0.01,\, 1/7, \, 0.5, \, 2)$, and the system is initialised with $(P_0, \, V_0, \, M_0) = (26,\, 20,\, 1)$; the drift $g$ is constant. Left: Phase portrait of the mispricing $\delta$ and the trend signal $M$ together with its nullclines and a sample trajectory. The streamlines' (blue) width and density encode the magnitude of the velocity field. Right: evolution of the price $P$, value $V$ and trend signal $M$.}
    \label{fig:spiral_determ}
\end{figure*}

\begin{figure*}[ht]
    \centering
    \includegraphics[width=0.7\linewidth]{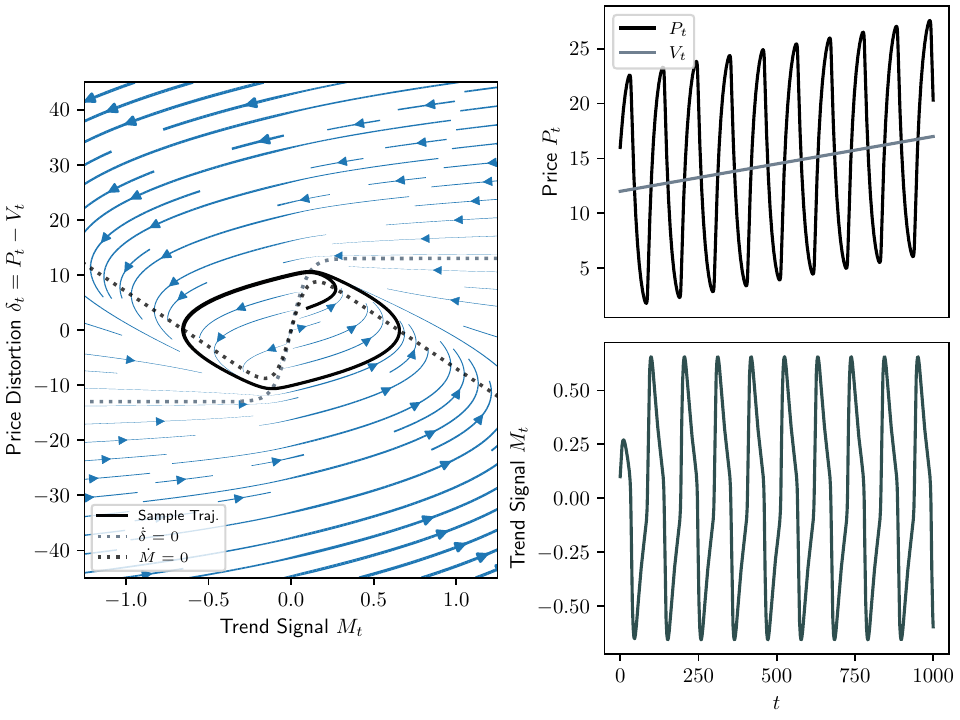}
    \caption{Same as Fig.~\ref{fig:spiral_determ} but in the case where the limit set is a limit cycle, $\kappa < \alpha (\beta\gamma -1)$. The parameters are $(\kappa, \, \alpha, \, \beta, \, \gamma) = (0.05,\, 1/7, \, 0.65, \, 10)$, and the system is initialised with $(P_0, \, V_0, \, M_0) = (16,\, 12,\, 0.1)$.}
    \label{fig:limitcycle_determ}
\end{figure*}

\subsubsection{Spiral Fixed Point}\label{sec:fixedpoint}

From the two nullclines, which intersect exactly once, it follows that there is a single fixed point (FP) at $(M^*, \delta^*)=(0, 0)$, the origin. The FP's stability type can be inferred by inspecting the Jacobian
\begin{align}
    J=
    \begin{pmatrix}
        -\kappa & \beta\gamma (1-\tanh^2 (\gamma M_t)) \\
        -\alpha\kappa & -\alpha + \alpha\beta\gamma (1-\tanh^2 (\gamma M_t))
    \end{pmatrix}
\end{align}
of system \eqref{eq:chiarella_on_mispricing} at the FP:
\begin{align}
    J|_{M=0, \delta=0}=J^*=
    \begin{pmatrix}
        -\kappa & \beta\gamma \\
        -\alpha\kappa & \alpha (\beta\gamma -1)
    \end{pmatrix}.
\end{align}
 $\det (J^*)=\alpha\kappa>0$ as the time scale $\alpha >0$, and mean reversion strength $\kappa>0$ for the dynamics not to diverge. This means that there are no saddles but only asymptotically (un)stable FPs. Further, tr$(J^*)=-\kappa + \alpha (\beta\gamma -1)$. The fixed point is stable when tr$(J^*)<$0, i.e. when $\kappa>\alpha (\beta\gamma -1)$. Consequently, tr$(J^*) -4\det (J^*) <0$, from which it follows that the FP is a \textit{spiral} (because the eigenvalues of $J^*$ then have non-zero imgaginary part). Hence, the bifurcation point at which the FP becomes unstable and the flow  in the $\delta$-$M$-plane changes qualitatively is
\begin{equation}
    \alpha^* = \frac{\kappa}{\beta\gamma -1}.
\end{equation}
That the fixed point is only stable when $\kappa > \alpha(\beta\gamma-1)$ shows that when value investors dominate trading, the deterministic part of the price dynamics converges to the fundamental value, where it remains forever. When chartists dominate trading, the FP becomes unstable and, in fact, a stable limit cycle emerges, such that there is a periodic motion of price around value. The emergence of the limit cycle is proven in the next section.

The condition $\kappa>\alpha (\beta\gamma -1)$ coincides with the result found in \cite{majewski2020co}, however, for their model the condition was only true for $g_t=0$ (which is not compatible with empirical results), while in our model the condition holds true generically.

\begin{figure*}[htbp]
    \centering
    \includegraphics[width=0.7\linewidth]{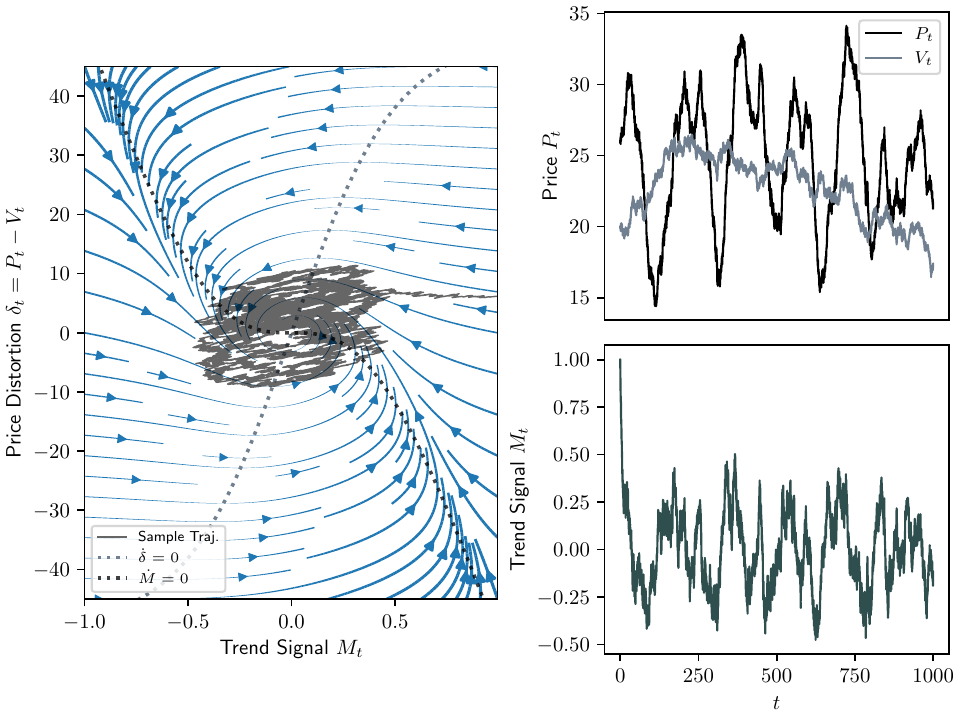}
    \caption{Same as Fig.~\ref{fig:spiral_determ} but in the presence of noise ($\sigma_N=0.35$ and $\sigma_V=0.2$).}
    \label{fig:spiral_stochatsic}
\end{figure*}

An example of such a stable spiral dynamics in the mispricing $\delta = P-V$, meaning that price converges to value, is depicted in Fig.~\ref{fig:spiral_determ} as a phase portrait in the $\delta$-$M$-plane alongside its price, value and trend signal trajectories in the deterministic case. Its stochastic analogue using the same parameters is given in Fig.~\ref{fig:spiral_stochatsic}.

\subsubsection{Hopf-Bifurcation: Emergence of a Limit Cycle}
In this section it will be shown that a stable limit cycle emerges when the FP loses its stability. This qualitative change of dynamics -- the loss of a FP's stability coinciding with the emergence of a periodic motion -- is known as a \textit{Hopf-bifurcation}. It occurs when a pair of complex conjugate eigenvalues of the Jacobian from the linearisation of the system around the FP crosses the imaginary axis in the complex plane as a parameter crosses its bifurcation point.

According to the Hopf-Bifurcation Theorem \cite{guckenheimer2013nonlinear, lorenz1993nonlinear}, two conditions on the eigenvalue pair of $J^*$, which in this case are
\begin{equation}
    \lambda_{1/2} = \frac{1}{2} \left( \alpha\beta\gamma -\alpha-\kappa \pm \sqrt{(-\alpha\beta\gamma +\alpha+\kappa)^2-4\alpha \kappa} \right),
\end{equation}
have to be fulfilled:
\begin{enumerate}
    \item The eigenvalue pair becomes purely imaginary at the bifurcation point:
    \begin{align*}
    \lambda_{1/2} (\alpha^*) &=
    \frac{1}{2} \left( \underbrace{\frac{\kappa\beta\gamma}{\beta\gamma -1}-\frac{\kappa}{\beta\gamma-1}-\kappa}_{=0} \pm \sqrt{-4\frac{\kappa^2}{\beta\gamma-1}} \right) \\
    &= i\kappa\sqrt{\frac{1}{\beta\gamma-1}},
    \end{align*}
    which is fulfilled since the condition involving tr$(J^*)$ in Sec. \ref{sec:fixedpoint} as $\beta\gamma >1$ is a necessary condition for the FP to become unstable. \\
    \item The real part of the derivative of the eigenvalues with respect to the bifurcation parameter, evaluated at the bifurcation point is non-zero:
    \begin{equation*}
    \frac{\partial\text{Re}(\lambda)}{\partial \alpha} (\alpha^*) = \beta\gamma -1 \neq 0,
    \end{equation*}
    which is is true for the same reason.
\end{enumerate}
The Hopf bifurcation may further be classified as \textit{supercritical}.

\begin{figure*}[htbp]
    \centering
    \includegraphics[width=0.7\linewidth]{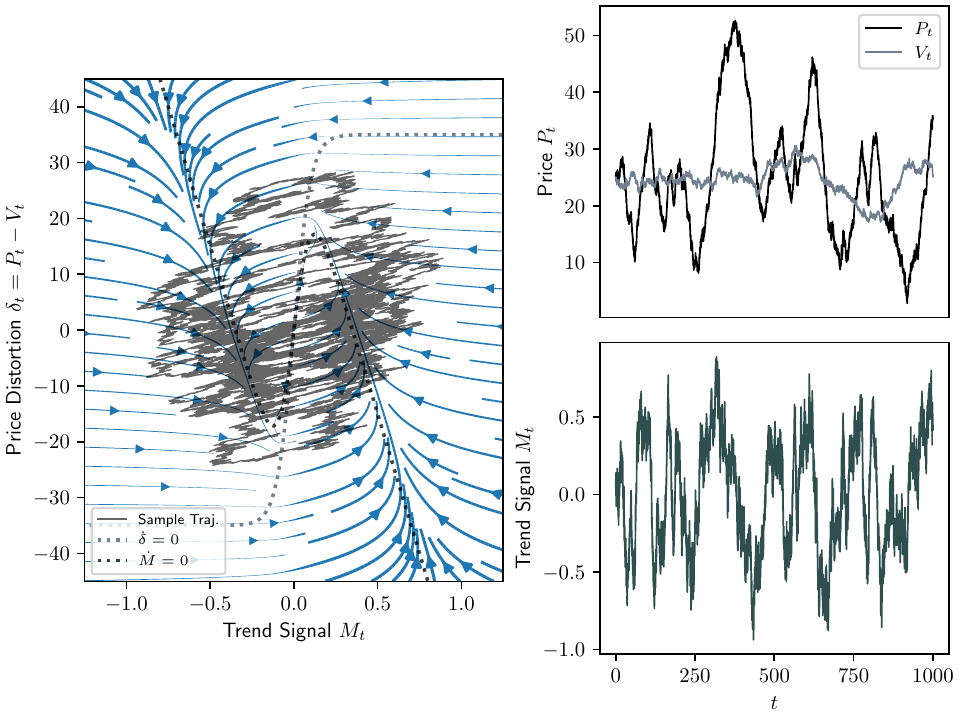}
    \caption{Same as Fig.~\ref{fig:limitcycle_determ} but in the presence of noise ($\sigma_N=0.6$ and $\sigma_V=0.2$). The parameters are $(\kappa, \, \alpha, \, \beta, \, \gamma) = (0.01,\, 1/7, \, 0.35, \, 10)$, and the system is initialised with $(P_0, \, V_0, \, M_0) = (26.5,\, 26,\, 0.1)$.}
    \label{fig:limitcycle_stochatsic}
\end{figure*}

An example of the deterministic dynamics (analog. to Fig.~\ref{fig:spiral_determ}) including the phase portrait with the limit cycle in the $\delta$-$M$-plane, as well as the evolutions of price, value and trend signal is provided with Fig.~\ref{fig:limitcycle_determ}. In Fig.~\ref{fig:limitcycle_stochatsic} is a full, stochastic example.

Thus, according to the model, price converges to value on long time scales if fundamentalists dominate trading, while it may oscillate when the presence of TFs is strong enough.

\section{Data} \label{sec:data}
All data used for the subsequent calibration of the model are {\it monthly} spot prices. In order to show the generality of the results, we study four different assset classes: Indices (IDX), Commodities (CMD), Bonds (BND) \& Currencies (FXR). In a subsequent paper, single name stocks will be further considered, which requires a model adaptation and modified calibration scheme, wherefore they are not included here.

To show that our model is an improvement over the one in \cite{majewski2020co} and to demonstrate the generality of the model, which can be applied to many asset classes, the model is calibrated on the same data set they used. The provider is \textit{Global Financial Data} and the data set available covers the period 1791 to 2015. As in \cite{majewski2020co}, we restrict the asset pool to only those products with a long enough history, which in the case of indices, bonds, and currencies means Australia, Canada, Germany, Japan, Switzerland, the UK and the US. Further, only exchange rates of the named countries' currencies against the US dollar are regarded. The considered commodities are copper, corn, crude oil, Henry Hub natural gas, live cattle, sugar, and wheat.

Furthermore, as in \cite{majewski2020co}, the time series of each asset is restricted to when it was publicly traded with high liquidity. Thus, the used exchange rate series start in 1973 as from 1945 to 1973 all considered currencies were pegged against the US dollar in the Bretton Woods system. Government bond prices are used only after 1920 for they were not liquid before. For all commodities the prices during World War II are excluded and additionally the period of 1939-1985 for crude oil. Equity index prices in particular display occasional extreme events as they are strongly impacted by political events, wherefore World War II is removed from the German and Japanese index' price series and World War I from the German and British ones. For Germany, the period around the so-called 'hyperinflation' of the Weimar Republic, which concerns the post WW I period is removed. Finally, the years 1973 and 1974, marked by the fall of the Bretton Woods system, are excluded from the British index. Whenever a time series is discontinued, the left end of the gap (with all the data preceding it) will be brought to the same level as the right end of the gap to avoid price jumps for which the model is not designed.

In addition to that, index prices are inflation adjusted by multiplying their nominal price with the respective Consumer Price Index (CPI) value belonging to that time stamp, normalised by (i.e. divided by) the final observed CPI value (such that the last price has a CPI multiplier of 1). Commodities are inflation adjusted with the US CPI in the same fashion.

Further details on the data set are provided in \cite{lemperiere2014two}.

\section{Calibration} \label{sec:calibration}

The two key steps in the calibration of a dynamical system of the type of~\eqref{eq: ModifiedChiarella} are a combination of the Expectation-Maximisation Algorithm and Kalman filtering, as done in \cite{majewski2020co} \footnote{For Bayesian filtering see, e.g., Särkkä \cite{sarkka2023bayesian}. The EM-Algorithm was introduced by Baum \cite{baum1970EMalgo} and extended to cases with incomplete observations by Dempster \cite{dempster1977maximum}. The joint usage of the EM-algorithm and filtering is due to Chen \cite{chen1981KalmanEM}.}. Following that paper, we explain the EM-Algorithm modified to fit model \eqref{eq: ModifiedChiarella} and other preliminary treatments of the data.

\begin{figure}[htbp]
    \centering
    %\includegraphics[width=\linewidth]{images/Dedrifting.pdf}
    % shown example is US index
    \includegraphics[width=\linewidth]{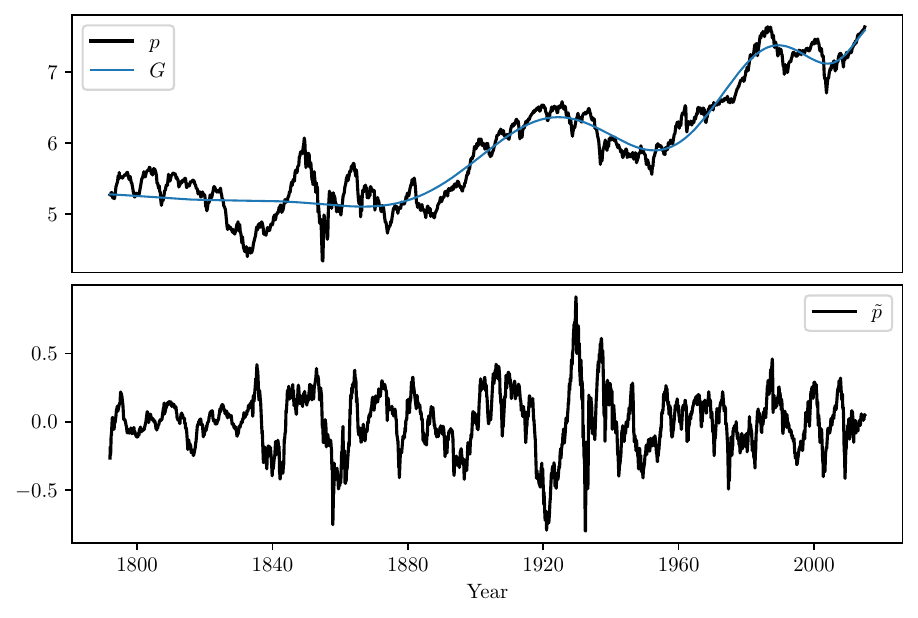}
    \caption{Evolution of the log-price $p$, the integrated drift $G$, and the de-drifted log-'price' $\tilde{p}$ of the US index. $G$ is estimated as a polynomial with one order per decade of data, i.e. a $22^\textup{nd}$ order polynomial here for data ranging from 1791-12 to 2014-12.}
    \label{fig:Dedrifting_example}
\end{figure}

A discrete time version of system~\eqref{eq: ModifiedChiarella}, where a time increment $\dd t=\Delta t=1$ corresponds to one month, is given by
\begin{align}
    \label{eq: chiarella_discretised}
    p_{t+1} &= p_t + \kappa (v_t - p_t) + \beta \tanh (\gamma m_t) + g_t + \eta^N_{t+1} \notag \\
    m_{t+1} &= (1-\alpha) m_t + \alpha (p_t-p_{t-1} - g_t) \\
    v_{t+1} &= v_t + g_t + \eta^V_{t+1}, \notag
\end{align}
where we use small cap notation for discrete time variables, and $\eta^{N/V}$ are Gaussian white noise processes with variance $\sigma^2_{N/V}$.
Due to the model set-up in Majewski et al. \cite{majewski2020co} and its implications on the Kalman relations and EM-algorithm the drift $g_t$ had to be fixed to a time independent value $g$. In our case, the (integrated) drift impacts price and value alike, our model becomes more canonical and $G_t= \int_0^t g_s \,\dd s$ may be removed from log-price series ex ante, allowing us to consider any time-dependent $g_t$. This improves the model as we do not generally find empirical evidence supporting the choice of a constant drift, and deem it too restrictive. Fig.~\ref{fig:Dedrifting_example} substantiates this claim for the US stock index whose evolution does not justify the assumption of a constant drift.

Denoting as $\tilde{p}$, $\Tilde{v}$, $\tilde{m}$ the de-drifted versions of $p$, $v$, and $m$, the formulation simplifies to
\begin{align}
    \label{eq:system_discretised_de-drifted}
    \tilde{p}_{t+1} &= \tilde{p}_t + \kappa (\tilde{v}_t - \tilde{p}_t) + \beta \tanh (\gamma \tilde{m}_t) + \eta^N_{t+1} \notag \\
    \tilde{m}_{t+1} &= (1-\alpha) \tilde{m}_t + \alpha (\tilde{p}_t-\tilde{p}_{t-1}) \\
    \tilde{v}_{t+1} &= \tilde{v}_t  + \eta^V_{t+1}. \notag
\end{align}

The question how the drift $g_t$ is to be chosen is quite important, since one can obviously find a perfect fit to the data by choosing $g_t \equiv p_{t+1} - p_t$, $\sigma_{N/V}=0$, $\beta=0$, such that $p_t = v_t$ at all times. This corresponds, in a sense, to the Efficient Market limit, where the price evolution is fully explained by changes of value. However, the well-documented presence of auto-correlation in the return time series (i.e. trend following on medium time scales and mean-reversion on long times scales \cite{black1986noise, asness2013value, lemperiere2014two, bouchaud2017black}) would mean that prices are {\it not} properly anticipated future values. 

Following the main tenet of the Chiarella model, we rather assume that trend-followers and noise traders have a non-zero impact on prices, i.e. $\beta > 0$ and $\sigma_N > 0$. We also assume that the long-term drift of value $g_t$ changes smoothly over time, higher frequency changes being captured by the noise term $\eta^V$. Assuming a business cycle of ten years, we propose to extract the long-term drift by fitting the price time series with a polynomial of order $k=\lfloor T/10 \rfloor$, where $T$ is the total length of the series in years.  However, we do not want $g_t$ to capture fluctuations on scales shorter than 5 years or so, since these should emerge from the dynamics of Chiarella's model itself. In fact, one should choose a time scale $T/k$ longer than $\kappa^{-1}$ in order to self-consistently assume that prices follow value on long enough time scales. Calibration will indeed suggest that $\kappa^{-1}$ is typically in the range $2$--$5$ years, see Table \ref{tab:4classes_Chiarella_lin}.  An example of log-price $p$, integrated drift $G$ and de-drifted price $\tilde{p}$ is shown in Fig.~\ref{fig:Dedrifting_example}, corresponding the US stock index, with an order 22 polynomial fit over circa 220 years.  We have checked that our results are very robust against changing $k$ in a reasonable range. A case study demonstrating this is given in Appendix~\ref{sec: dedrift_change_robustness} where the values of calibrated parameters are given for $k=14, 22$  and $30$, on the example of the US stock index.

\subsection{EM-Algorithm } \label{sec:EM_algo}

The estimation of the parameters and the de-drifted log-value methodology overlaps with the algorithm proposed in \cite{majewski2020co}.
Thus, we only briefly describe the algorithm here as it coincides with the one of Ref. \cite{majewski2020co}, but now with $g=0$ since $\tilde{p}$, $\tilde{m}$, $\tilde{v}$ in Eqs.~\eqref{eq:system_discretised_de-drifted} are already de-drifted. Further, we propose methods different from those in \cite{majewski2020co} for the ex ante estimation of the trend parameters $\gamma$ and $\alpha$, which are then fixed in the EM-algorithm to reduce the number of parameter estimates from six to four and because there exist simpler and more canonical ways of estimating them than through the EM-algorithm.

Note that were the parameters known, the unobservable fundamental value could be inferred as what is known in the control theory literature as a hidden or latent variable via Bayesian filtering techniques \cite{sarkka2023bayesian}. Indeed, system \eqref{eq: ModifiedChiarella} is linear in $v$ and the noise is assumed to be Gaussian, hence the optimal filter is a Kalman filter. If this was not the case, particle filters could be used.

Having obtained a first value proxy by initialising the calibration, the EM-algorithm is used to obtain a set of optimal parameters based on the current fundamental value by maximising an otherwise difficult to compute marginal log-likelihood by instead maximising a joint log-likelihood. Each iteration follows a two step procedure:
\begin{enumerate}
    \item \textbf{E-step}: calculation of a conditional expectation of the joint log-likelihood of the posterior distribution over the hidden variable $v$, given past prices and the current best guess of the parameters.
    \item \textbf{M-step}: calculation of the parameters by optimising the joint log-likelihood.
\end{enumerate}
After each iteration there is a new estimate of the fundamental value $v$ together with the set of parameters $\theta = (\kappa, \beta, \sigma_N, \sigma_V, v_0)$, until the algorithm terminates when the increase in likelihood falls below a tolerance of $\epsilon=10^{-5}$.

For the model with a non-linear demand function of the fundamentalists, system~\eqref{eq: ModifiedChiarellaNonlinear}, we also use the adapted version ($g=0$) of the algorithm presented in \cite{majewski2020co}, utilising unscented Kalman filtering to treat the cubic fundamentalists' demand.

\subsection{Estimation of $\alpha$ and $\gamma$}

The trend time scale $\alpha$ is chosen as the time scale that maximises the Sharpe ratio of the assets' {\it de-drifted} trend signal $\tilde{m}$. Choosing $\tilde{m}$ over $m$ also undercuts the appropriate criticism of trend signals often used in the literature that are defined on past returns directly, thus reflecting mostly the long-only bias (buying and holding an asset while its price and value tend to increase over long horizons due to overall market growth), rather than the actual short-term to medium-term trend. The Sharpe ratio is the expected return from a strategy in excess of a benchmark return (here: the return from the long-only strategy) divided by the standard deviation of that excess return:
\begin{equation}
    \label{eq:Sharpe}
    SR = \frac{\mathbb{E} [\tilde{r}]}{\sqrt{Var [\tilde{r}]}},
\end{equation}
where $\tilde{r}$ is the excess (log-)return of the signal. This quantifies the expected performance of an investment after adjusting for its involved risk. To wit,
\begin{equation}
    \alpha = \underset{\alpha'}{\text{argmax}} \,\, SR (\tilde{m}(\alpha')).
\end{equation}
The typical EWMA time scale of the trend signals $\tilde{m}$ that maximises the Sharpe ratio is $\alpha \approx 1/5$.
\footnote{$\alpha$ could be dissected further for each individual time series but we refrain from doing so as it varies as much over the centuries as over the products. Further, we find (as in \cite{majewski2020co}) that results are almost invariant for $\alpha \in \{1/4, \, 1/5, \,1/6, \,1/7\}$. The sloppiness analysis in Sec.~\ref{sec:sloppy} will formalise this justification.}

Next, it is shown why the fundamentalists' demand imbalances are chosen to be a hyperbolic tangent of the trend signal. Note, albeit, that the hyperbolic tangent is just an example of a function that saturates for relatively large values of the signal. The necessary conditions that such a function should obey have been derived for the original Chiarella model in Ref. \cite{chiarella1992dynamics}.

\begin{figure}[htbp]
    \centering
    \includegraphics[width=\linewidth]{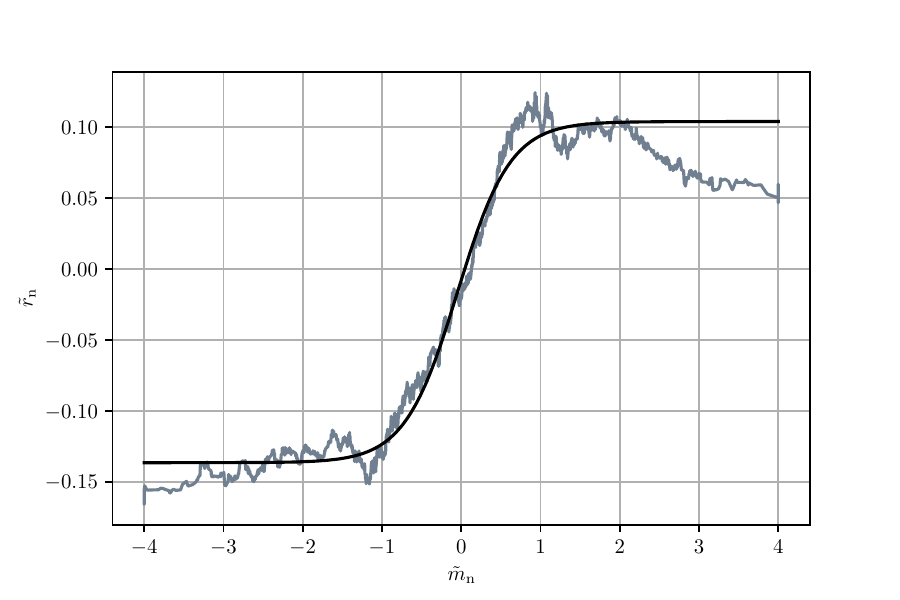}
    \caption{De-drifted and normalised future returns $\tilde{r}_\mathrm{n}$ as a function of the trend signal $\tilde{m}_\mathrm{n}$ calculated from all returns from currencies against the US dollar. The grey curve is a rolling average over 1000 consecutive points along the abscissa of the real data aggregated over all six FXR pairs, while the solid black line is a parametric fit of eq.~\eqref{eq:tanh} over all data.}
    \label{fig:tanh_FXR}
\end{figure}

This functional relationship relating the (normalised) returns $\tilde{r}_\mathrm{n}$ and the trend signal $\tilde{m}_\mathrm{n}$ calculated from those returns is depicted in Fig.~\ref{fig:tanh_FXR}. Its shape is common to all regarded asset classes, serving an ex-post justification of Chiarella's assumptions.
Fig.~\ref{fig:tanh_FXR} shows a slight departure from the hyperbolic tangent for large positive trend signals beyond two standard deviations. This phenomenon has been reported in the literature as trend-reversion \cite{bouchaud2017black,schmidhuber2021trend_reversion}. However, practitioners usually clip their signals at $\pm 2 \sigma$, eliminating this effect. Because such strong trend signals are rare anyway, we deem the $\tanh$ an appropriate parametric choice.
The returns and trend signals had to be normalised in Fig.~\ref{fig:tanh_FXR} in order to make them comparable among different assets that may show different levels of volatility and thus differently sized returns.

In order to estimate the parameter $\gamma$ per instrument, the function
\begin{equation}
\label{eq:tanh}
    h(x) = a + b\tanh (\tilde{\gamma} x + c)
\end{equation}
is fitted to data from the sets of assets belonging to one asset class. For each product the slope $\gamma$ is then brought back into its natural units via
\begin{equation}
    \gamma = \frac{\tilde{\gamma}}{\sqrt{Var [\tilde{m}]}},
\end{equation}
$Var [\tilde{m}]$ is the variance of the trend signal. This is done for each product.

The $\gamma$-values for each product are reported together with all calibration output in Table~\ref{tab:4classes_Chiarella_lin} in Appendix~\ref{app:calib_results}.

\subsection{Calibration Results \& Excess Volatility} \label{sec:calib_res_nonstock}
In order to improve the robustness of the calibration and because the algorithm generally exhibits difficulties in pinning down $\sigma_V$, a multi-step calibration is used.
The three step calibration can be summarised as follows:
\begin{enumerate}
    \item Calibration of model~\eqref{eq:system_discretised_de-drifted} on the de-drifted log-price series $\tilde{p}$ of each asset to get a first estimate of the parameters (after ex-ante estimations of $G$, $\alpha$, $\gamma$).\\
    \item Calibration of the factor $\Sigma = \langle \frac{\sigma_N}{\sigma_V} \rangle_{a \in \mathcal{A}}$ per asset class $\mathcal{A}$ by maximising the cumulated likelihood of all the assets $a$ in one class (keeping all other parameters fixed).\\
    \item Recalibration of model~\eqref{eq:system_discretised_de-drifted} (or non-linear model: Eqs.~\eqref{eq: ModifiedChiarellaNonlinear}) per asset but with $\sigma_V = \frac{1}{\Sigma} \sigma_N$ fixed.
\end{enumerate}

\begin{figure*}[ht]
    \centering
    \includegraphics[width=0.8\linewidth]{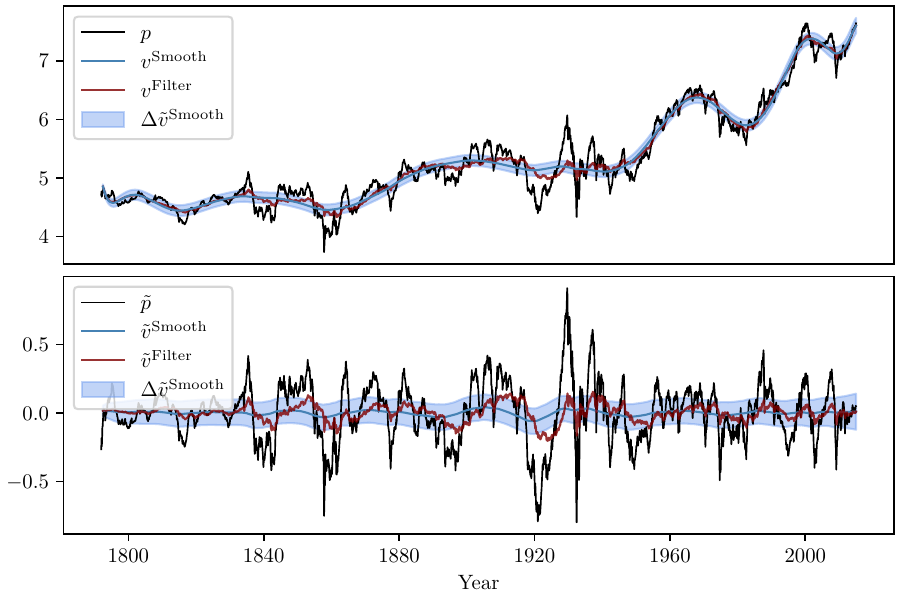}
    \caption{Top: Evolution of the log-price $p$ together with the filtered and smoothed fundamental values $v^\text{Filter}$ and $v^\text{Smooth}$ for the US stock index. $\Delta v^\text{Smooth}$ denotes the confidence interval obtained as one standard deviation of the value from the Kalman relations. Bottom: same for the de-drifted log-price $\tilde{p}$ and values.}
    \label{fig:US_calib}
\end{figure*}

\begin{figure*}[ht]
    \centering
    \includegraphics[width=0.8\linewidth]{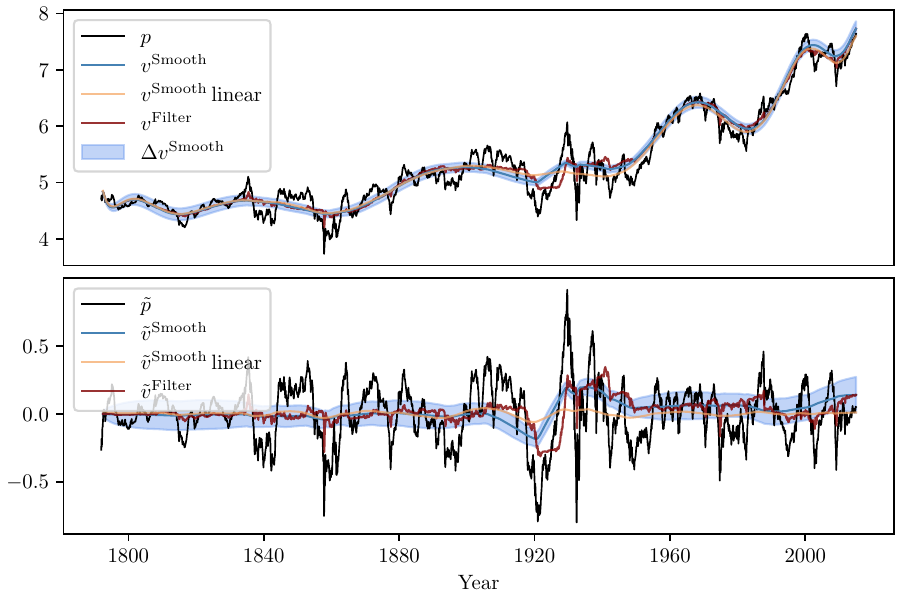}
    \caption{Same as Fig.~\ref{fig:US_calib} but for the non-linear (cubic) model. The smoothed value from the linear model is given for comparison (orange).}
    \label{fig:US_calib_nonlin}
\end{figure*}

From Table~\ref{tab:asset_factors} (center column), which lists the calibrated factor $\Sigma$ between the two noise sources per asset class, we infer that the contribution from noise traders is crucial and much exceeds the noise of the fundamental (log-)value process $v$.
\begin{table}[htbp]
    \centering
    \renewcommand{\arraystretch}{1.1} % Increase row height by 1.5 times
    \begin{tabular}{l|l|l} % `ll` denotes two left-aligned columns
        \toprule
        \textbf{Asset Class} & $\Sigma$ (linear) & $\Sigma$ (non-linear) \\ % Headers
        \midrule
        Indices & $3.87 \, \pm\, 0.61$ & $3.81 \pm 0.65$\\
         Commodities & $3.62 \, \pm\, 1.62$ & $3.35 \pm 1.13$\\
         Currencies & $5.93 \, \pm\, 1.10$ & $5.89 \pm 1.09$\\
         Bonds & $13.94\pm 4.46$ & $13.82 \pm 4.35$ \\
        \bottomrule
    \end{tabular}
    \caption{Estimated ratio $\Sigma = \frac{\sigma_N}{\sigma_V}$ between the two noise sources per asset class. The error is given by the standard deviation within one class. Left: linear model, right: non-linear. The value of $\Sigma$ is found to be nearly identical for the two models. Furthermore, replacing $\sigma_N$ by $\sigma_P$ (price volatility) gives almost indistinguishable results.}
    \label{tab:asset_factors}
\end{table}

This may be interpreted as a quantification of the famous excess volatility puzzle, first formalized by Shiller in 1981 \cite{shiller1981stock}; see, e.g., also \cite{leroy2006excess}. Within our framework, this excess volatility is mostly due to excess trading from noise traders, which has been reported on all asset classes studied here \cite{shiller1981stock, deaton1992behaviour, meese1983empirical, campbell1991yield}. Trend-following activity, on the other hand, does not contribute much to short-term volatility because the signal is computed over rather long time scales. However, trend-following is responsible for further long term decoupling between price and value.  

More precisely, we find (comp. Table~\ref{tab:asset_factors}) that the volatility of the price due to noise traders is significantly larger (by a factor 4 to 14) than the volatility of the fundamental value in all cases. This is one of the central result of our study, and justifies the title of this paper. In general, $\sigma_N$ is indistinguishable from $\sigma_p$, the price volatility calculated via the dedrifted log-returns, within the error margins, justifying its usage in determining the excess volatility.

For indices and commodities, $\sigma_N$ is typically around four times as large as $\sigma_V$, suggesting that prices depart from value due to strong excess trading. Such an amplification factor is compatible with other estimates from the literature, see e.g. \cite{wehrli2022excess}.

For exchange rates, the ratio between the noise trader volatility and fundamental volatility is higher, with a value around $6$, twice as large as the value reported in \cite{wehrli2022excess}. The largest ratio is found for bonds, for which we report a ratio of almost $12$. However, such a high ratio for bonds is not due to an extreme amount of excess trading on that asset class, which would correspond to a large $\sigma_N$ that we do not observe, see Table.~\ref{tab:4classes_Chiarella_lin}. Instead, it stems from the fundamental volatility being particularly small for bonds, as expected since the fundamental value of bonds is expected to be much more stable than the fundamental value of indices. Foreign exchange rates are in this sense intermediate.

As an illustration, the log-price $p$ together with its filtered and smoothed calibrated fundamental values $v$ are shown in Fig.~\ref{fig:US_calib} (top plot). The blue shaded area indicates one standard deviation of the smoothed value according to the Kalman smoother relations. The bottom plot of Fig.~\ref{fig:US_calib} provides the same insight but on the de-drifted (log-)price $\tilde{p}$ and value $\tilde{v}$.

In addition to that, the calibrated parameters of the linear model in Table~\ref{tab:4classes_Chiarella_lin} are such that none of the assets satisfies the bifurcation condition for oscillations as for all assets $\beta \gamma< 1$, ensuring that $\kappa > \alpha (\beta\gamma -1)$. Similar results are reported in \cite{majewski2020co}. It was further shown in Chiarella et al. \cite{chiarella2008stochastic_bifurc} that the limit set being a cycle is a necessary condition for the distribution of trend signals to be bimodal. Since we {\it do} find that for some assets the distribution of mispricing is bimodal (even with $\beta\gamma<1$, see Table \ref{tab:bimodality_check}) we are compelled to reject the linear specification of the Chiarella model and turn to the non-linear version, see Eqs. \eqref{eq: ModifiedChiarellaNonlinear}, which has a much richer phase diagram that is in fact not yet fully explored analytically. 

The calibration results for the non-linear model using the unscented Kalman filter with the cubic fundamentalist demand are detailed in Table~\ref{tab:4classes_Chiarella_nonlin} in Appendix ~\ref{app:calib_results}. As for the linear model, the results are illustrated on the US stock index, which is illustrated in Fig.~\ref{fig:US_calib_nonlin} -- see Appendix~\ref{app:calib_results}, in which the estimated filtered and smoothed fundamental values from the cubic model are shown with an error bar alongside the de-drifted log-price and the linear comparison. As was also noted in \cite{majewski2020co}, we find a small, often negative value of $\kappa$ (enhancing trend-like behaviour for small mispricing) and a stauchly positive value of $\kappa_3$, confirming that mean-reverting behaviour only becomes appreciable for large mispricings.
The noise ratios $\Sigma$ are listed in the right column of Tab.~\ref{tab:asset_factors}; they are very similar to those for the linear model, which means that $\sigma_N$ is estimated similarly in both models (note that for the non-linear model $\sigma_V$ is fixed to the value obtained from the linear model).

In conclusion of this section, we have shown that it is now possible to calibrate meaningfully our modified versions of Chiarella's  model, both linear and non-linear, on {\it individual} time series, while it was previously only possible to jointly calibrate classes of similar time series \cite{majewski2020co}.

\section{Mispricing Distribution \& Bimodality} \label{sec:mispricing_dist}

\begin{table*}
    \centering
     \rowcolors{2}{gray!20}{white}
    \begin{tabular}{l|c|c|c||c||c}
        \toprule
    \textbf{} & \textbf{Filtered} & \textbf{Smoothed} & \textbf{Bimodality} & \textbf{Bimodality} & \textbf{J-S} \\
    \cellcolor{white} & \cellcolor{white} & \cellcolor{white} & \cellcolor{white}\textbf{(Empirical)} & \cellcolor{white}\textbf{(Numerical)} & \cellcolor{white}\textbf{Distance} \\
    \hline
    \textbf{US} & 0.364 & 0.353 & \ding{55} & $\checkmark$ & 0.128 \\%0.188 \\
    \textbf{UK} & 0.011 & 0.010 & $\checkmark$ & $\checkmark$ & 0.136 \\%0.169 \\
    \textbf{AU} & 0.001 & 0.001 & $\checkmark$ & $\checkmark$ & 0.164 \\%0.178 \\
    \textbf{CH} & 0.031 & 0.098 & \ding{55} & $\checkmark$ & 0.152 \\%0.164 \\
    \textbf{JP} & 0.001 & 0.001 & $\checkmark$ & $\checkmark$ & 0.202 \\%0.165 \\
    \textbf{CA} & 0.067 & 0.182 & \ding{55} & $\checkmark$ & 0.148 \\%0.212 \\
    \textbf{DE} & 0.019 & 0.047 & $\bm{\backsim}$ & $\checkmark$ & 0.200 \\%0.224 \\
    \hline
    \textbf{SUGAR} & 0.007 & 0.001 & $\checkmark$ & $\checkmark$ & 0.304 \\%0.380 \\
    \textbf{CORN} & 0.002 & 0.001 & $\checkmark$ & $\checkmark$ & 0.172 \\%0.228 \\
    \textbf{LCATTLE} & 0.639 & 0.898 & \ding{55} & \ding{55} & 0.115 \\%0.136 \\
    \textbf{WHEAT} & 0.706 & 0.258 & \ding{55} & $\checkmark$ & 0.166 \\%0.196 \\
    \textbf{COPPER} & 0.099 & 0.040 & \ding{55} & $\checkmark$ & 0.257 \\%0.270 \\
    \textbf{NATGAS} & 0.074 & 0.313 & \ding{55} & $\checkmark$ & 0.245 \\%0.304 \\
    \textbf{CRUDE} & 0.001 & 0.785 & $\bm{\backsim}$ & \ding{55} & 0.214 \\%0.237 \\
    \hline
    \textbf{USBND} & 0.362 & 0.346 & \ding{55} & \ding{55} & 0.149 \\%0.167 \\
    \textbf{UKBND} & 0.023 & 0.014 & $\bm{\backsim}$ & \ding{55} & 0.136 \\%0.157 \\
    \textbf{CHBND} & 0.233 & 0.026 & \ding{55} & $\checkmark$ & 0.224 \\%0.245 \\
    \textbf{JPBND} & 0.002 & 0.001 & $\checkmark$ & $\checkmark$ & 0.203 \\%0.276 \\
    \textbf{AUBND} & 0.331 & 0.306 & \ding{55} & \ding{55} & 0.158 \\%0.152 \\
    \textbf{CABND} & 0.322 & 0.341 & \ding{55} & \ding{55} & 0.123 \\%0.145 \\
    \textbf{DEBND} & 0.004 & 0.073 & $\bm{\backsim}$ & $\checkmark$ & 0.148 \\%0.144 \\
    \hline
    \textbf{CHFUSD} & 0.077 & 0.277 & \ding{55} & $\checkmark$ & 0.247 \\%0.357 \\
    \textbf{JPYUSD} & 0.175 & 0.275 & \ding{55} & \ding{55} & 0.138 \\%0.168 \\
    \textbf{AUDUSD} & 0.738 & 0.117 & \ding{55} & $\checkmark$ & 0.168 \\%0.148 \\
    \textbf{GBPUSD} & 0.890 & 0.643 & \ding{55} & \ding{55} & 0.151 \\%0.169 \\
    \textbf{CADUSD} & 0.855 & 0.452 & \ding{55} & \ding{55} & 0.133 \\%0.153 \\
    \textbf{EURUSD} & 0.024 & 0.165 & \ding{55} & $\checkmark$ & 0.270 \\%0.304 \\
    \hline
    \end{tabular}
    
    \caption{p-values of the Silverman test for bimodality (column 2 and 3). The null hypothesis is unimodality with a significance level of $0.02$. The test is on the mispricing $\delta$, once with the empirical filtered fundamental value and once with the smoothed. \ding{55}: acceptance of the unimodal null hypothesis for both empirical time series, \checkmark: twofold rejection (bimodality), $\bm{\backsim}$: inconclusiveness. Center Right (Numerical): modality test results of simulated time series of the non-linear model, using parameters from Table~\ref{tab:4classes_Chiarella_nonlin}. Right: Jensen-Shannon (J-S) distances between the empirical and numerical mispricing distributions. A J-S distance of $0.2$ corresponds to a Kolmogorov-Smirnov distance of $8 \%$ (i.e. the maximum distance between two cumulative distributions functions). For comparison, the J-S distance between LCATTLE and AU is $0.19$.}
    \label{tab:bimodality_check}
\end{table*}

The asset price dynamics described by Eqs.~\eqref{eq: ModifiedChiarella} may be further classified by the shape of the resulting mispricing distribution $\rho (\delta)$ (where again $\delta = p-v$), in particular with respect to its uni- or bimodality. 

For the classical Chiarella model (without drift) this was numerically investigated by Chiarella et al. in \cite{chiarella2008stochastic_bifurc}, where the conditions for $\rho$ to be bimodal were established. Empirically, bimodal distributions of price distortions independent of Chiarella-type models were reported in \cite{schmitt2017bimodality}. Majewski et al. confirmed this finding for a modified Chiarella model with a constant drift in the fundamental value $v$ \cite{majewski2020co}; see also \cite{lux2021can} for further discussions.  

As alluded to above, the Chiarella model can only generate bimodal mispricing distributions if $\kappa < \alpha (\beta\gamma -1)$, that is when the limit set is a stable limit cycle \cite{chiarella2008stochastic_bifurc}. It was then shown in Majewski et al. that in the latter case, the distribution of the trend signal is also bimodal \cite{majewski2020co}. Still empirical results are at odds with these predictions: not only one finds bimodality of mispricing even when the calibrated parameters are such that $\beta \gamma < 1$, but one does not necessarily find a corresponding bimodality in the trend distribution. The non-linear version of the model does not suffer from these limitations. The mispricing distribution results presented in this section are restricted to the model~\eqref{eq: ModifiedChiarellaNonlinear} with a cubic demand function, where bimodality can occur without a bimodal trend signal distribution when $\kappa_3 > 0$ and $\kappa \leq 0$. 

For the analytical stationary probability density is unknown, we probe bimodality via Silverman's test for multimodality \cite{silverman1981bimodal_test}, which tests for a distribution having a minimum of $k+1$ modes, while the null hypothesis is a distribution with at most $k$ modes. Consequently, we perform the test with $k=1$, such that a rejection of the null hypothesis is tantamount to rejecting a unimodal distribution, hence suggesting bimodality. More than two modes is not possible within the models investigated here, and there is no such empirical evidence either. A significance level of $0.02$ is chosen; this means the null-hypothesis of unimodality is rejected when the p-value is below 0.02.

Silverman's test is performed on two different kinds of empirical mispricing series $\delta$ for the simple reason that the Kalman relations allow for two different notions of fundamental value $v$. The first is the filtered value, which is determined through the dynamical system and the information of \textit{past} prices and values. The second is the smoothed value, which takes both \textit{past and future} information into account. As a result, the filtered value is the value that could have been known to the fundamentalist at the time of trading, while the smoothed value is an ex post best estimate of what the true fundamental value really was.

Those results are summarised in the first three columns of Table~\ref{tab:bimodality_check}, which focus on the empirical mispricing distributions, i.e. those from the real log-prices $p$ and the two types of calibrated log-values $v$. If both types of empirical mispricings suggest bimodality within the given significance, we mark the series as empirically bimodal ($\checkmark$). If none of the two reject the null-hypothesis of unimodality, the asset is classified as empirically unimodal (\ding{55}). And if one accepts and one rejects the null-hypothesis, the test is marked as inconclusive ($\bm{\backsim}$). 

Table~\ref{tab:bimodality_check} shows that there is in many cases clear empirical evidence for a rejection of unimodality. At the 2\% significance level, the empirical mispricing distributions of stock indices are bimodal in nearly half of the cases. For commodities there is further empirical evidence for bimodality in the cases of sugar and corn. In addition to that, the test is inconclusive for Crude Oil, UK Bonds and German Bonds.
\begin{figure*}[ht]
    \centering
    \includegraphics[width=0.7\linewidth]{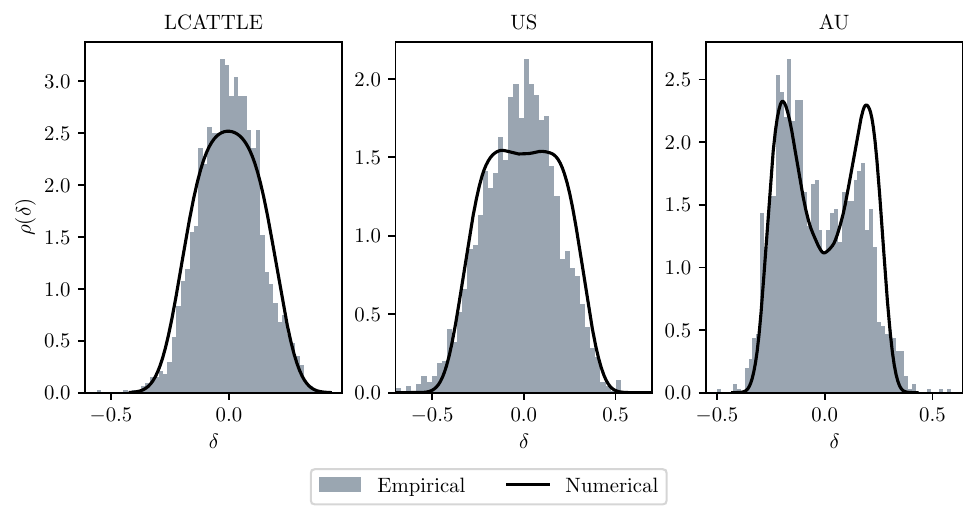}
    \caption{Empirical: mispricing distribution of the log-price and the smoothed value derived from the Kalman filter with parameters from Table~\ref{tab:4classes_Chiarella_nonlin} for the commodity live cattle (LCATTLE) and the US and Australian (AU) stock index. Numerical: mispricing distribution from a simulation of system~\eqref{eq: ModifiedChiarellaNonlinear} with $T=10^5$, d$t=0.01$ and parameters from Table~\ref{tab:4classes_Chiarella_nonlin}.}
    \label{fig:mispricing_dists_spots}
\end{figure*}

It is however known \cite{mammen1992bimod} that the Silverman test suffers from a conservatism bias -- especially for small samples like we have here -- where the null-hypothesis is falsely not rejected. Therefore, we have repeated the test on data generated from numerical simulations of the non-linear model~\eqref{eq: ModifiedChiarellaNonlinear} using the calibrated optimal parameters listed in Tab.~\eqref{tab:4classes_Chiarella_nonlin}. Simulating the model for a total duration of $T=10^5$ with time increments of d$t=0.01$, yields $N=10^7$ data points per product and alleviates the data scarcity problem. Indeed, the test on simulated data yields very accurate results even for weak bimodalities, i.e. when the system being close to the critical bifurcation point. The results are given in the second rightmost column of Table~\ref{tab:bimodality_check}. It shows that in all the cases where the test suggests bimodality based on the empirical data, it finds bimodality based on the numerical data, confirming the results as well as the success of the calibration from a different angle. Furthermore, in the cases where the test on empirical data is inconclusive, the test on numerical data leans towards bimodality (except for one asset, crude oil). The numerical study also finds bimodality in some cases where the empirical one did not.
Strikingly, the numerical test suggests bimodality for all indices, for most commodities and around half of the bonds and currencies. Note that our flexible definition of the long-term drift, which allows for low-frequency oscillations akin to ``business cycles'', tends to lessen any sign of bimodality.

Hence, bimodality appears to be the rule rather than the exception. This is interesting for two reasons: first, it suggests that in many cases assets are more often under- or overpriced than correctly priced. According to numerical data this holds true for all considered indices -- a stunning albeit not necessarily surprising result, and in line with the result reported in \cite{schmitt2017bimodality} for the S\&P 500, and in \cite{majewski2020co} for the US and Canadian stock markets. Second, it provides evidence that in such cases prices perform noisy oscillations around value, at least according to the studied model, and even after having accounted for business cycles.

One example of an empirical and its respective numerical mispricing distribution in the case where the distribution is unimodal (live cattle) and another one in the bimodal case (Australian stock index) is given in Fig.~\ref{fig:mispricing_dists_spots} alongside one of the cases (US stock index) where the empirical and numerical distributions suggest a different type of modality (even though the bimodality in the numerical data is extremely weak)\footnote{Schmitt et al. \cite{schmitt2017bimodality} report a more pronounced bimodality based on monthly historic S\&P 500 prices. This may be due to differences in estimation of fundamental value where they apply Shiller's method based on discounting the index' real dividend payments \cite{shiller2000irrational}, instead of our flexible definition of a time varying drift with low-frequency oscillations. It might also be due to their overestimation the trend signal, which is defined on log-returns directly, instead of returns in \textit{excess} of some benchmark, such as the long-term drift (as done in our study), or the sectorial or market drifts as often done in practice, esp. for stocks. The extreme bimodalities obtained from simulating several models based on the trend signals of \cite{schmitt2017bimodality} may be a symptom of this.}. The comparison of the empirical and numerical distributions show that the mispricings can be captured rather well, considering that non-linear model~\eqref{eq: ModifiedChiarellaNonlinear} is a highly schematic model with only three investor types. More importantly, the calibration relies on price trajectories and is unaware of the empirical mispricing distribution, so a good match with the predicted mispricing distribution can be seen as an independent validation of the model. Jensen-Shannon (J-S) distances between the empirical and numerical distributions are given in the rightmost column of Tab.~\ref{tab:bimodality_check} to quantify the similarity between numerical and empirical distributions for all assets \cite{endres2003new}.

The J-S distance is the square root of the J-S divergence, which is a symmetrised and smoothed version of the well-known Kullback-Leibler divergence. It has the benefit of being a metric that can be interpreted as a distance measure. The distance's lower bound of zero means identical distributions, while the upper bound indicating maximal difference is one \cite{endres2003new}. As the J-S distance requires the same domain support and binning for both distributions, the domain is chosen to be between the minimum and maximum of the empirical and numerical distribution with the number of bins being the square root of the length of the empirical time series. This is important as the length difference between both series is many orders of magnitude, such that using the length of the shorter series is critical to have an empirical distribution that is not zero for many intervals in the domain.

Before calculating the J-S distances, we have made sure that the mean and the variance of the empirical distributions exactly match those of the numerical simulation. This is only approximately true with the parameters obtained from the calibration, but a small shift of these parameters in the direction of the gradient of the variance allows us to fix this issue with a minimal change of the log-likelihood of the calibration (less than $5 \%$ in most cases).

\section{Sloppiness Analysis} \label{sec:sloppy}

The so-called \textit{sloppiness analysis} is based on the Fisher information matrix, with the aim of gaining some insight about the hierarchy of parameter importance in a model or in a dynamical system, see \cite{gutenkunst2007sloppy} and \cite{quinn2022information} for more recent developments. In this context, a parameter (or combination of parameters) is termed \textit{sloppy} if a perturbation in its direction does not change the model output significantly: the model is insensitive to the exact value of that parameter, further implying that it is hard to estimate empirically -- parameter uncertainty is high, as encapsulated by the Cramer-Rao bound. Likewise, a \textit{stiff} parameter (or combination of parameters) leads to significant changes in a model's output upon perturbation, implying that it can be estimated well on empirical data. 

The crux of the method lies in calculating the eigen-decomposition of the Hessian matrix of a loss function $\mathcal{L}$, where the loss is a quantifier of the change in model output upon perturbation of the set of model parameters $\theta$ to $\theta' = \theta(1+\Delta)$, where $\Delta$ is small. Here, the loss function is defined as the (normalized) $L_2$ distance between the realizations of observable $y$ measured before and after a perturbation is applied:
\begin{equation}
\mathcal{L} (\theta, \theta') = \frac{1}{T}  \sum_{t =1}^T \left( \frac{y_t(\theta)-y_{t}(\theta')}{\sigma} \right)^2,
    \label{eq:loss_sloppiness}
\end{equation}
where $t$ are increments of the simulation time $T$.
Note that the random seed must be fixed in this analysis to only measure the loss due to the parameter perturbation and not due to noise. Further, the beginning of each time series may want to be dropped due to stationarity. $\sigma$ is the standard deviation over time of $y_{t}(\theta)$. In the following we will choose the mispricing $\delta$ as the observable $y$.

The model sensitivity to parameter variations may then be regarded through the Hessian of the loss, a.k.a. the Fisher information matrix:
\begin{equation}
    \mathcal{H}_{ij} = \frac{\dd^2 \mathcal{L}(\theta,  \theta')}{\dd\log (\theta_i)\, \dd\log (\theta_j)} \Big|_{\theta'=\theta},
    \label{eq:Hessian_sloppiness}
\end{equation}
where it is standard to take log-derivatives to regard relative parameter changes as parameters usually have inconsistent units and their magnitudes may span multiple decades.  Subsequent analysis of $\mathcal{H}$ via its eigendecomposition corresponds to an approximation of the surfaces of constant model deviations as $N$-dimensional ellipsoids \cite{gutenkunst2007sloppy}. Numerically, $\mathcal{H}_{ij}$ can be computed using only first derivatives of $y_t(\theta)$, see e.g. \cite{naumann2024exploration}.

\begin{figure}
    \centering
    \includegraphics[width=\linewidth]{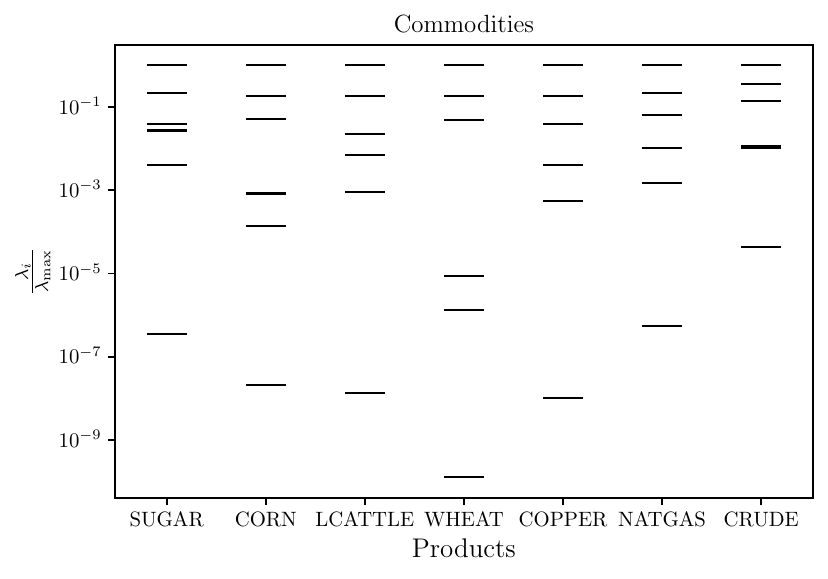}
    \caption{Parameter sensitivity spectrum due to the eigenvalues of eq.~\eqref{eq:Hessian_sloppiness} for the linear model~\eqref{eq:system_discretised_de-drifted} and the asset class commodities. Per asset the spectrum is rescaled by its maximum eigenvalue for visualisation.}
    \label{fig:sloppy_eigvals__mispricing_CMD}
\end{figure}

A model is termed sloppy if its sensitivity eigenvalue spectrum spans multiple decades in a rather consistent manner, in other words that the eigenvalues decay very quickly with rank, meaning that only very few parameters (or linear combinations thereof) can be identified. Applying this rationale to the mispricing $\delta = p-v$ in the linear model~\ref{eq:system_discretised_de-drifted} over $N=6$ parameters $\theta = (\kappa,\, \beta,\, \gamma,\, \alpha,\, \sigma_N,\, \sigma_V)$ with optimal parameters from Table~\ref{tab:4classes_Chiarella_lin} and $\Delta = 10^{-2}$, strong evidence for sloppiness is reported as the sensitivity spectra span from five to nine decades (comp. Appendix ~\ref{app:sloppy}, Fig.~\ref{fig:sloppy_linear_eigvals_spot_all}). The same is reported for the non-linear model with the additional parameter $\kappa_3$  (Fig.~\ref{fig:sloppy_NONlinear_eigvals_spot_all} in Appendix ~\ref{app:sloppy}). For the linear model and the asset class commodities, this is depicted in Fig.~\ref{fig:sloppy_eigvals__mispricing_CMD} exemplarily.

Next, the eigenvectors of the Hessian in eq.~\eqref{eq:Hessian_sloppiness} for the linear model are analysed. Since it is not sensible to regard each time series' eigenvectors standalone and because financial time series within one asset class share many key characteristics, we analyse the \textit{average} Hessian within one asset class.

The eigenvectors of the Hessian point in the eigendirections in parameter space. Thus, one can infer from those whether the eigendirections coincide with individual parameter axes, or whether parameter \textit{combinations} determine a system's dynamics and how it reacts to perturbations. Since the Hessian of the loss of the system perturbed from its optimal parameters $\theta$ is determined, those eigendirections (descendingly with the magnitude of the corresponding eigenvalue) denote the direction in which the fit is degraded the quickest, i.e. it orders the directions towards which the model dynamics are most sensitive.

\begin{figure}[htbp]
    \centering
    \includegraphics[width=\linewidth]{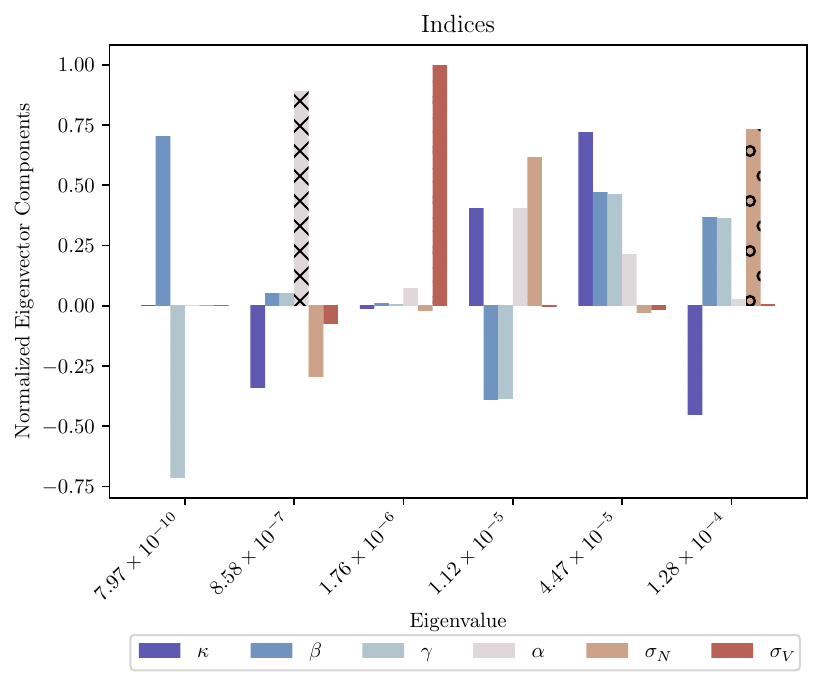}
    \caption{The 6 normalised eigenvectors for stock indices, ranked by eigenvalue magnitude (in increasing order) of the average Hessian for the linear model. The observable in the loss $\mathcal{L}$ (comp. eq.~\ref{eq:loss_sloppiness}) is the price distortion $\delta = p-v$. The total number of observations is $N=10^4$.}
    \label{fig:sloppy_eigvecs_lin_IDX}
\end{figure}

For stock indices, the normalised eigenvectors of the Hessian $\mathcal{H}$ of the linear model perturbed around its optimal parameters (comp. Table~\ref{tab:4classes_Chiarella_lin}) and ranked by their eigenvalues are given in Fig.~\ref{fig:sloppy_eigvecs_lin_IDX}.
Interestingly, the eigenvectors for other asset classes (commodities, currencies, bonds) are qualitatively very similar compared to Fig.~\ref{fig:sloppy_eigvecs_lin_IDX}. Overall it is clear at first glance that the eigendirections in parameter space do not generally align with the parameter directions because most modes are mixed, except for $\sigma_V$, which has its own isolated mode for all four asset classes (mode four, corresponding to the fourth largest eigenvalue, in all classes except currencies, comp. Fig.~\ref{fig:sloppy_eigvecs_lin_IDX}).
Moreover, the eigenvectors of all asset classes show that the contribution of $\beta$ and $\gamma$ to the eigendirections is always the same. This is no surprise considering that the local linearisation of the model as well as the bifurcation condition depend solely on the product $\beta\gamma$ and not on the parameters individually (comp. Sec.~\ref{sec:stability_analysis}). More generally, the different eigenvectors (or {\it modes}) can be interpreted in the context of the model.

\begin{itemize}
    \item The first mode reflects the quick degradation in quality of fit when $(\sigma_N,\, \beta,\, \gamma)$ are perturbed in one direction and $\kappa$ in the other. The first three parameters control the dispersion between price and value. In particular, when they are increased (decreased), the dispersion increases (decreases) as $\sigma_N$ is the largest influence on the variance of the price and both $\beta$ and $\gamma$ are parameters associated with trend following, also increasing the departure of price from value when increased. Meanwhile, $\kappa$ has the opposite effect, describing mean reversion towards value. Hence, decreasing (increasing) $\kappa$ also results in larger dispersion from value. The first mode may thus be termed the \textit{variance mode}. In total, it can be concluded that this mode governs the variance of the price distortion; it is also the stiffest parameter direction in the system to which it is most sensitive and which can be calibrated most reliably (which is confirmed, e.g., by the small errors on $\sigma_N$ and $\kappa$ in Appendix ~\ref{app:calib_results}). In all other asset classes the first mode looks qualitatively similar. Quantitatively it can additionally be deduced through comparison of the first modes in Fig.~\ref{fig:sloppy_linear_eigvals_spot_all} that for currencies the contributions of the parameters associated with trend following, $\beta$ and $\gamma$, are almost negligible, which suggests relatively weak trend following in currency markets, compatible with the results of \cite{lemperiere2014two} but not really born out by the analysis of \cite{hurst2017century}. For commodities the opposite is visible: $\beta$ and $\gamma$ -- and thus trend following in general -- are the most pronounced, a further known fact for after all almost the entire CTA industry was built on this strategy \cite{elaut2019trends}.
\item 
The second mode in Fig.~\ref{fig:sloppy_eigvecs_lin_IDX} may be termed the \textit{critical} or \textit{bifurcation mode} as it depends on the parameters occurring in the bifurcation condition (comp. Sec.~\ref{sec:stability_analysis}). The mode shows that an increase (decrease) in mean reversion strength (by $\kappa$) accompanied by a simultaneous increase (decrease) in trend following (through $\beta \gamma$ and $\alpha$) leads to a deterioration of the fit even though there is no direct implication on the level of the price distortion as the effects counterbalance each other. The bifurcation condition -- and therewith the overall dynamical state -- is however sensitive to such perturbations as the different parameters do not enter the condition equally but in a non-linear way.
\item The third mode is similar to the first mode in composition, except for a stronger contribution of $\alpha$. As in the first mode, $\sigma_N$ is the dominant contributor. $\kappa$ seems to be a larger contributor in this mode in all asset classes than on mode one. The relative sign of $\kappa$ and $\beta \gamma$ are inverted compared to the first mode, as a consequence of the orthogonality condition acting on the different eigenvectors.
\item The fourth mode, the \textit{value noise mode}, is the only case in which a parameter direction coincides precisely with an eigendirection. It describes the response of the price distortions to perturbations in $\sigma_V$. Its eigenvalue is however two orders of magnitude lower than that of the first mode. This implies that one needs to perturb the system $\sqrt{\lambda_1 / \lambda_4} \approx 10$ times as hard in that direction to achieve a comparable variation in model output. This, of course, implies that $\sigma_V$ is relatively loosely constrained, which makes its estimation ten times harder. At the same time it means that the exact value of $\sigma_V$ is less relevant compared to $\sigma_N$, $\kappa$, $\beta$ and $\gamma$.
\item The fifth mode can be interpreted as the \textit{trend speed mode} as it is mostly determined by $\alpha$. Its small associated eigenvalue confirms our previous statement that changes in $\alpha$ (that we have hard-coded to $1/5$) do not change the results significantly, neither qualitatively, nor quantitatively (comp. the footnote in Sec.~\ref{sec:calibration}).
\item The sixth and last mode is a pure trend following mode as it only depends on the trend parameters $\beta$ and $\gamma$. It is the only mode in which the two have different orientations. This mode can be interpreted as the \textit{trend saturation mode}. It is the consequence of the breaking of the linear dependence of the trend signal on $\beta \gamma$ as higher order terms of the $\tanh$ function become relevant. This mode has the smallest impact as the sloppiest direction for only returns beyond two standard deviations fall in this saturation regime, which is rare, such that the overall influence is comparatively small. Principally, it shows how an increase (decrease) in $\beta$ accompanied by a decrease (increase) in $\gamma$ reduces the quality of fit, which is an immediate consequence of the $\tanh$ and its contributions that are $O (\beta \gamma^3)$ and ensures that $\beta$ and $\gamma$ can both be estimated and not just their product  -- albeit not with very high precision.
\end{itemize}
The non-linear model, Eqs.~\eqref{eq: ModifiedChiarellaNonlinear}, whose sloppy analysis is detailed in Appendix~\ref{app:sloppy}, naturally has one more eigendirection as it has an additional parameter, $\kappa_3$. As it turns out, the mode interpretation is very similar to the linear model. The addiotional seventh eigendirection can be interpreted as a \textit{value mode}, quantifying the response to perturbations in $(\kappa, \, \kappa_3)$, and ranks as the sixth mode, whereas the sixth from the linear model becomes the final seventh mode.

%%%%%%%%%%%%%%%%%%%%%%%%%%%%%%%%%%%%%%%%%%

\section{Conclusion}\label{sec:conclusion}
In this paper the dynamical interplay between trend and value anomalies that pervade (almost) all financial markets was revisited. Specifically, the generalized Chiarella model proposed by Majewski et al. \cite{majewski2020co} was corrected for its analytical shortcomings, which also impacted its calibration. We proposed a new self-consistent model, in which the stability conditions for the dynamically possible phases exactly match the model dynamics -- even for non-zero, arbitrary long term value drift. This was achieved by letting the trend signal and price dynamics, and not only the fundamental value, be drift-dependent. The idea is to define the trend signal on \textit{mispricing} returns rather than on standard returns, removing any long term bias, which should not be considered as part of the trend. 

Our model is therefore able to accommodate arbitrary time-dependent drifts, whereas previously only linear drifts were allowed, which we deem unsatisfactory. A calibration scheme adapted to this new model was proposed and implemented. This leads to a notable estimation improvement as it enables one to calibrate the model on individual price time series, whereas previously only asset-class-wide calibrations were possible. As in the literature, this was performed on a model that is linear in the fundamentalist's demand as well as one that is non-linear (cubic). We find that only the non-linear model is consistent with many of the stylised facts, including the bimodality of the mispricing distribution while the trend signal remains unimodal.

One important output of the calibration is the fundamental value of an asset. By proposing a new price vs. value variance estimation technique, the ratio between the noise trader induced volatility and the value volatility could be estimated per asset class. This ratio sheds light on the long-standing excess volatility puzzle as it confirms and quantifies by how much price volatilities are amplified over volatilities in value and that changes in value do not justify the amplitude of price changes whatsoever, putting the rationality and efficiency of prices into question. Our estimate of an amplification of a factor 4 for stock indices is comparable to other estimates from the literature, including Shiller's original paper \cite{shiller1981stock}. Differences in excess volatility per asset class could be qualitatively accounted for. It was possible to separate the variance contribution of noise traders and of fundamental value for each asset individually.

Besides, the distribution of instantaneous mispricings was empirically and numerically analysed for the non-linear model. Statistical tests confirm the existence of bimodality, which have been previously reported \cite{schmitt2017bimodality, majewski2020co}. In fact much stronger evidence was found than in Majewski at al. \cite{majewski2020co}, especially for stock indices and commodities, while for bonds and currencies less so. This finding shakes the Efficient Market Hypothesis to its core for it suggests that assets are more likely to be mispriced than correctly priced -- but it also pinpoints markets (bonds and currencies) that are closer to efficiency.

A Hessian or sloppy analysis allowed for a systematic multi-parameter sensitivity study of the model to small changes in parameter combinations, defining a strong hierarchy in the eigenvalues of the Fisher information matrix. This analysis, on the one hand, justifies why certain parameters are difficult to estimate, but on the other hand also suggests that their exact value may not be crucial to the model's dynamical signature. It may also be interesting from a regulatory standpoint as it can quantify which market contributors and effects can affect the price dynamics most notably. We were able to explain these parameter directions in the context of market perturbation modes.

This research opens the path to several follow-up questions and topics: first, it would be interesting to see how the parameters have evolved over centuries. This may be a difficult endeavour as the dynamical time-scales are on the order of decades, making sequential calibrations difficult in many cases due to data scarcity. However, it could elucidate whether prices and markets have become more efficient with time and whether levels of mispricing have grown or shrunk, see also \cite{schmidhuber2021trend_reversion}. Second, an extension of the model towards single stocks would be worthwhile in order to understand their level of mispricing for different economic sectors, and to study the excess volatility puzzle further through our lens and in its original context. We will tackle this question in a subsequent publication. It would also be interesting to repeat this analysis on crypto assets, which do not have a tangible notion of fundamental value. Finally, a more micro-founded model of demand leads to an enhanced version of the Chiarella model that we are currently investigating.

\section*{Acknowledgements}
We would like to thank Julius Bonart, Damien Challet and Doyne Farmer for their insights and useful discussions, as well as Fabian Aguirre-Lopez for mathematical remarks.
This research was conducted within the Econophysics
\& Complex Systems Research Chair, under the aegis of
the Fondation du Risque, the Fondation de l’Ecole polytechnique, the Ecole polytechnique and Capital Fund
Management.

\clearpage

\bibliographystyle{vancouver}
\bibliography{biblio}

\clearpage

\appendix

%\makeatletter
%\renewcommand{\p@subsection}{}
%\renewcommand{\p@subsubsection}{}
%\makeatother
%\renewcommand{\thesection}{\arabic{section}}
%\renewcommand{\thesubsection}{\arabic{subsection}}
%\renewcommand{\thesubsubsection}{\arabic{subsubsection}}

\small 
\onecolumngrid

\begin{center}
\Large
Appendix to the Paper \textit{Revisiting the Excess Volatility Puzzle Through the Lens of the Chiarella Model}
\end{center}
Authors: Jutta G. Kurth, Adam A. Majewski and Jean-Phlippe Bouchaud

\noindent
Journal: PLOS ONE

\section{Calibration Results}\label{app:calib_results}

Here, the calibration outputs for the spots (indices, commodities, futures, bonds) described in Sec.~\ref{sec:data} are presented.

\subsection{Linear model}

The calibration parameters for the linear model of the calibration step 3 in Sec.~\ref{sec:calib_res_nonstock} are given in Table~\ref{tab:4classes_Chiarella_lin}. Their standard errors, which are obtained through a Hessian analysis of the likelihood function under small perturbations in the case of $\kappa$, $\beta$, $\sigma_N$ and $v_0$ are listed in Table~\ref{tab:calib_errors_4classes_lin}. In the case of $\gamma$ the standard error is obtained similarly through the fit of the hyperbolic tangent (comp. Eq.~\eqref{eq:tanh} and Fig.~\ref{fig:tanh_FXR}).

\begin{table}[ht]
    \centering
    \rowcolors{2}{white}{gray!20}
    % Define a command to add padding
    %\newcommand{\mystrut}{\rule{0pt}{2.25ex}} % Adjust as needed for height
    % Set table width using p{width}
    %\begin{tabular}{>{\mystrut}l|>{\centering\arraybackslash}p{1.3cm}|>{\centering\arraybackslash}p{1.3cm}|>{\centering\arraybackslash}p{1.3cm}|>{\centering\arraybackslash}p{1.3cm}|>{\centering\arraybackslash}p{1.3cm}|>{\centering\arraybackslash}p{1.3cm}|>{\centering\arraybackslash}p{1.3cm}}
    %\begin{tabular}{>{\mystrut}l|*{7}{C{1.3cm}}}
    %\begin{tabular}{l|*{7}{C{1.3cm}}}
    \begin{tabular}{l|c|c|c|c|c|c|c}
        %\toprule
        {} &  $\kappa$ &   $\beta$ &  $\gamma$ &  $\sigma_N$ &  $\sigma_V$ & $v_0$ & $\bar{\mathcal{L}}$ \\
        \hline
        \textbf{US}      &  0.027 &  0.076 &  4.168 &    0.043 &    0.011 &  0.017 & 1.723 \\
        \textbf{UK}      &  0.026 &  0.100 &  4.101 &    0.036 &    0.009 &  0.019 & 1.907 \\
        \textbf{AU}      &  0.028 &  0.079 &  4.238 &    0.039 &    0.010 &  0.012 & 1.823 \\
        \textbf{CH}      &  0.034 &  0.100 &  4.232 &    0.043 &    0.011 &  0.024 & 1.708 \\
        \textbf{JP}      &  0.028 &  0.095 &  4.147 &    0.058 &    0.015 &  0.128 & 1.398 \\
        \textbf{CA}      &  0.045 &  0.101 &  4.138 &    0.045 &    0.012 &  0.027 & 1.678 \\
        \textbf{DE}      &  0.036 &  0.125 &  4.279 &    0.044 &    0.011 &  0.025 & 1.702 \\ 
        \hline
        \textbf{SUGAR}   &  0.063 &  0.146 &  2.246 &    0.073 &    0.020 & -0.011 & 1.186 \\
        \textbf{CORN}    &  0.096 & -0.027 &  2.324 &    0.119 &    0.033 &  0.103 & 0.699 \\
        \textbf{LCATTLE} &  0.270 &  0.381 &  2.198 &    0.045 &    0.013 &  0.004 & 1.276 \\
        \textbf{WHEAT}   &  0.081 & -0.002 &  2.518 &    0.092 &    0.025 & -0.066 & 0.960 \\
        \textbf{COPPER}  &  0.055 &  0.059 &  2.100 &    0.061 &    0.017 &  0.044 & 1.365 \\
        \textbf{NATGAS}  &  0.195 & -0.111 &  2.321 &    0.174 &    0.048 &  0.104 & 0.300 \\
        \textbf{CRUDE}   &  0.089 &  0.276 &  2.766 &    0.106 &    0.029 & -0.362 & 0.812 \\ 
        \hline
        \textbf{USBND}   &  0.069 &  0.048 &  5.316 &    0.046 &    0.003 & -0.002 & 1.644 \\
        \textbf{UKBND}   &  0.084 &  0.068 &  5.382 &    0.048 &    0.003 &  0.006 & 1.599 \\
        \textbf{CHBND}   &  0.062 &  0.089 &  4.802 &    0.051 &    0.004 & -0.004 & 1.546 \\
        \textbf{JPBND}   &  0.038 & -0.008 &  6.066 &    0.087 &    0.006 & -0.035 & 1.009 \\
        \textbf{AUBND}   &  0.066 &  0.067 &  5.849 &    0.045 &    0.003 & -0.001 & 1.645 \\
        \textbf{CABND}   &  0.068 &  0.064 &  4.901 &    0.036 &    0.003 & -0.000 & 1.896 \\
        \textbf{DEBND}   &  0.066 &  0.103 &  4.574 &    0.045 &    0.003 & -0.005 & 1.668 \\ 
        \hline
        \textbf{CHFUSD}  &  0.035 &  0.032 &  6.455 &    0.035 &    0.006 & -0.007 & 1.923 \\
        \textbf{JPYUSD}  &  0.033 &  0.057 &  6.451 &    0.032 &    0.005 & -0.017 & 1.996 \\
        \textbf{AUDUSD}  &  0.040 &  0.028 &  7.011 &    0.033 &    0.005 & -0.002 & 1.980 \\
        \textbf{GBPUSD}  &  0.044 &  0.051 &  6.701 &    0.029 &    0.005 & -0.013 & 2.102 \\
        \textbf{CADUSD}  &  0.019 &  0.004 &  6.764 &    0.019 &    0.003 & -0.013 & 2.510 \\
        \textbf{EURUSD}  &  0.025 &  0.030 &  6.403 &    0.032 &    0.005 & -0.014 & 2.000 \\
        %\bottomrule
    \end{tabular}
    \caption{Calibration results for the fit of the de-drifted linear model~\eqref{eq:system_discretised_de-drifted} on the four asset classes (calibration step 3. in Sec.~\ref{sec:calib_res_nonstock}). 
|    First group: stock indices, second: commodities, third: bonds, fourth: currencies. The data for some of the assets, esp. commodities, goes back 200 years. $\bar{\mathcal{L}}$ is the predictive log-likelihood normalised by the series length. Std. errors for the values are listed in Table~\ref{tab:calib_errors_4classes_lin}. $\beta\gamma<1$ for all assets.}
    \label{tab:4classes_Chiarella_lin}
\end{table}

\begin{table}[ht]
    \centering
    \rowcolors{2}{white}{gray!20}
    % Define a command to add padding
    %\newcommand{\mystrut}{\rule{0pt}{2.25ex}} % Adjust as needed for height
    % Set table width using p{width}
    %\begin{tabular}{>{\mystrut}l|>{\centering\arraybackslash}p{1.3cm}|>{\centering\arraybackslash}p{1.3cm}|>{\centering\arraybackslash}p{1.3cm}|>{\centering\arraybackslash}p{1.3cm}|>{\centering\arraybackslash}p{1.3cm}}
    \begin{tabular}{l|c|c|c|c|c}
        %\toprule
        {} &  $\Delta \kappa$ &   $\Delta \beta$ &  $\Delta \gamma$ & $\Delta\sigma_N$ & $\Delta v_0$  \\
        \hline

\textbf{US}      &  0.007 &  0.023 & 1.121 & 0.001 &   0.011 \\
\textbf{UK}      &  0.005 &  0.028 & 1.103 & 0.002 &   0.041 \\
\textbf{AU}      &  0.009 &  0.025 & 1.140 & 0.003 &   0.008 \\
\textbf{CH}      &  0.028 &  0.295 & 1.139 & 0.002 &   0.427 \\
\textbf{JP}      &  0.008 &  0.030 & 1.116 & 0.002 &   0.195 \\
\textbf{CA}      &  0.012 &  0.027 & 1.113 & 0.002 &   0.127 \\
\textbf{DE}      &  0.007 &  0.022 & 1.151 & 0.001 &   0.067 \\ \hline
\textbf{SUGAR}   &  0.015 &  0.047 & 4.700 & 0.002 &   0.004 \\
\textbf{CORN}    &  0.054 &  0.146 & 4.864 & 0.017 &   0.773 \\
\textbf{LCATTLE} &  0.217 &  0.364 & 4.599 & 0.002 &   0.005 \\
\textbf{WHEAT}   &  0.021 &  0.055 & 5.270 & 0.003 &   0.168 \\
\textbf{COPPER}  &  0.010 &  0.058 & 4.394 & 0.003 &   0.178 \\
\textbf{NATGAS}  &  0.130 &  0.254 & 4.858 & 0.012 &   0.176 \\
\textbf{CRUDE}   &  0.019 &  0.050 & 5.788 & 0.004 &   0.423 \\ \hline
\textbf{USBND}   &  0.014 &  0.031 & 3.746 & 0.002 &   0.000 \\
\textbf{UKBND}   &  0.022 &  0.038 & 3.792 & 0.002 &   0.005 \\
\textbf{CHBND}   &  0.019 &  0.038 & 3.384 & 0.004 &   0.001 \\
\textbf{JPBND}   &  0.028 &  0.031 & 4.274 & 0.007 &   0.148 \\
\textbf{AUBND}   &  0.020 &  0.034 & 4.122 & 0.002 &   0.000 \\
\textbf{CABND}   &  0.016 &  0.032 & 3.453 & 0.002 &   0.000 \\
\textbf{DEBND}   &  0.114 &  0.120 & 3.223 & 0.003 &   0.065 \\ \hline
\textbf{CHFUSD}  &  0.014 &  0.026 & 5.566 & 0.001 &   0.013 \\
\textbf{JPYUSD}  &  0.013 &  0.023 & 5.562 & 0.001 &   0.196 \\
\textbf{AUDUSD}  &  0.014 &  0.029 & 6.045 & 0.001 &   0.001 \\
\textbf{GBPUSD}  &  0.015 &  0.023 & 5.777 & 0.001 &   0.085 \\
\textbf{CADUSD}  &  0.044 &  0.025 & 5.832 & 0.001 &   0.144 \\
\textbf{EURUSD}  &  0.090 &  0.088 & 5.521 & 0.001 &   0.102 \\
\end{tabular}
\caption{Standard errors for the estimated parameters of Table~\ref{tab:4classes_Chiarella_lin}.}
\label{tab:calib_errors_4classes_lin}
\end{table}
The error of $\sigma_V$ is not listed in Table~\ref{tab:calib_errors_4classes_lin} as $\sigma_V$ is fixed via $\sigma_V = \frac{1}{\Sigma} \sigma_N$ as detailed in Sec.~\ref{sec:calib_res_nonstock}. Through the errors of $\Sigma$ and $\sigma_N$, it can principally be approximated via Gaussian error propagation:
\begin{align}
    \Delta \sigma_V \approx \sqrt{\left(\frac{\partial \sigma_V}{\partial \Sigma} \Delta\Sigma \right)^2 + \left(\frac{\partial\sigma_V}{\partial\sigma_N} \Delta\sigma_N\right)^2},
    \label{eq: error_sig_v}
\end{align}
where $\Delta\sigma_N$ is provided in Table~\ref{tab:calib_errors_4classes_lin} and $\Delta\Sigma$ in Table~\ref{tab:asset_factors}.

We realise that four of the assets report negative $\beta$ in Table~\ref{tab:4classes_Chiarella_lin}, while $\beta >0$ according to the model as it would otherwise be another driver of mean-reversion rather than trending. This problem may, however, be due to the calibration and not real as for all cases $\beta$ may as well be positive within the error margins provided by Table~\ref{tab:calib_errors_4classes_lin}. Further, the normalised likelihood $\bar{\mathcal{L}}$ of the model is always lowest for those assets with negative $\beta$ within their respective group, affirming possible calibration irregularities.

\begin{figure}[htbp]
    \centering
    \includegraphics[width=0.56\linewidth]{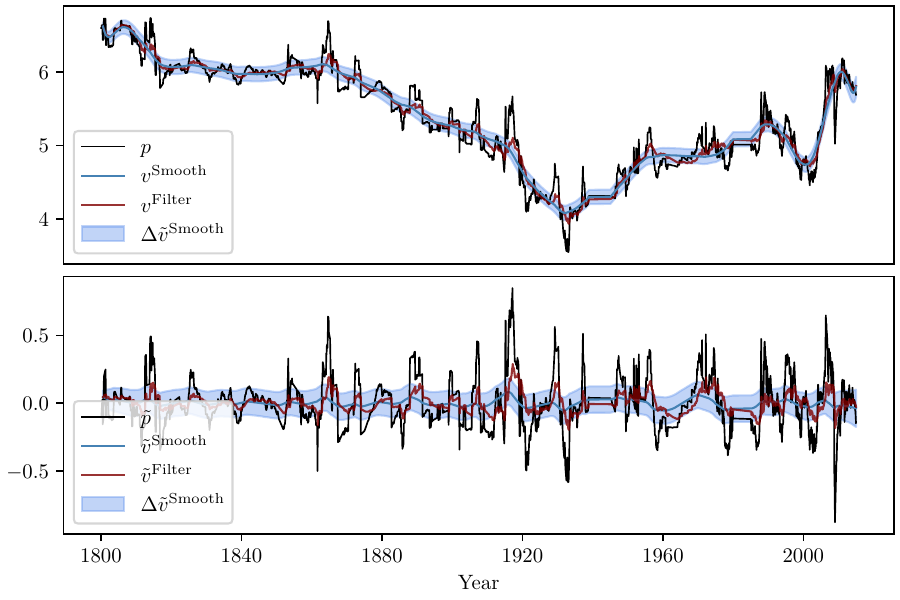}
    \caption{Same as Fig.~\ref{fig:US_calib} but for the evolution of the price of the commodity copper.}
    \label{fig:Copper_calib}
\end{figure}

\begin{figure}[htbp]
    \centering
    \includegraphics[width=0.56\linewidth]{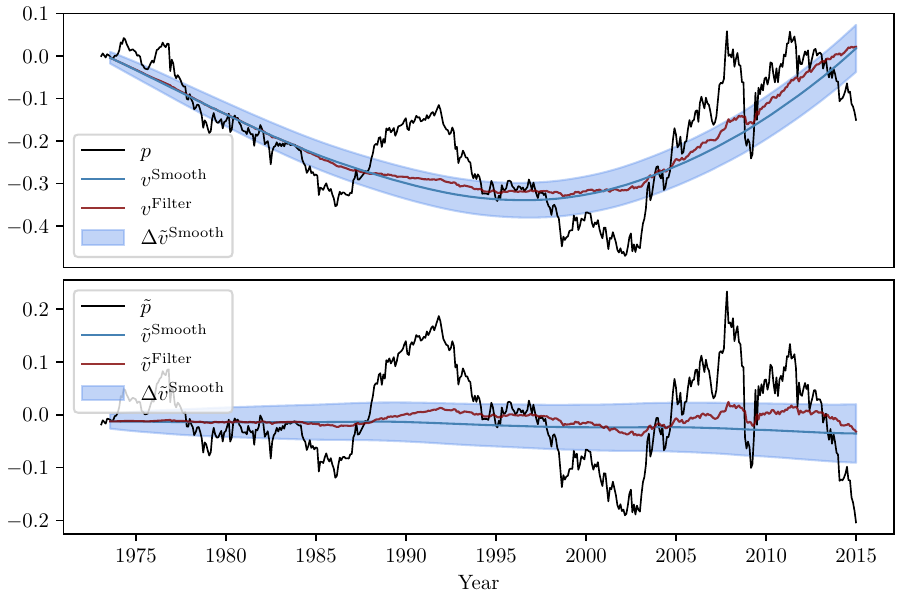}
    \caption{Same as Fig.~\ref{fig:US_calib} but for the Canadian dollar vs. US dollar currency pair.}
    \label{fig:CADUSD_calib}
\end{figure}

\begin{figure}[htbp]
    \centering
    \includegraphics[width=0.56\linewidth]{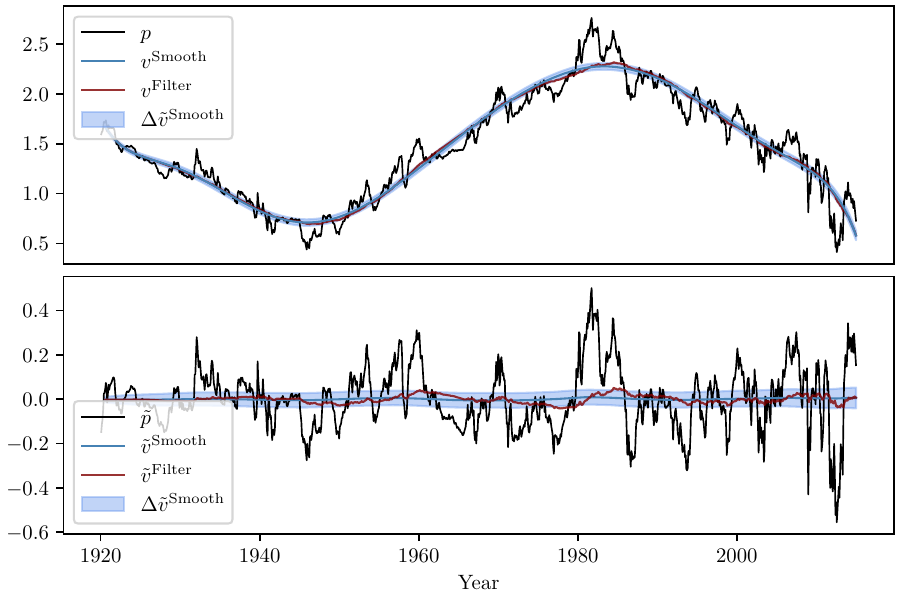}
    \caption{Same as Fig.~\ref{fig:US_calib} but for the US government bond.}
    \label{fig:USBND_calib}
\end{figure}

In the following, we provide further graphic examples of calibrated fundamental values according to the output summarised in Table~\ref{tab:4classes_Chiarella_lin}; in particular Copper for commodities (Fig.~\ref{fig:Copper_calib}), the Canadian dollar vs. the US dollar for currency pairs (Fig.~\ref{fig:CADUSD_calib}) and the US bond for the bond asset class (Fig.~\ref{fig:USBND_calib}).

\subsection{Non-linear model}

The non-linear (cubic) model ~\eqref{eq: ModifiedChiarellaNonlinear} can be calibrated using the algorithm detailed in \cite{majewski2020co} using unscented Kalman filtering to treat the non-linearity in (log-)values. Note that as described in Sec.~\ref{sec:calibration} for the linear model, the time series considered here must first be dedrifted; then their algorithm may be directly deployed with $g=0$ (analog. to the linear model). Furthermore, as the algorithm has problems pinning down $\sigma_V$, we fix it, as mentioned in calibration step 3. in Sec.~\ref{sec:calib_res_nonstock}, except using a dedrifted and discretised version of model~\eqref{eq: ModifiedChiarellaNonlinear} instead of its linear counterpart, Eqs.~\eqref{eq:system_discretised_de-drifted}.

The analogue of Table~\ref{tab:calib_errors_4classes_lin} for the de-drifted non-linear version of model~\eqref{eq: ModifiedChiarellaNonlinear}, listing all the calibrated parameters, is given in Table~\ref{tab:4classes_Chiarella_nonlin}. Table~\ref{tab:calib_errors_4classes_nonlin} lists the standard errors of those estimates. $\Delta \sigma_V$ could again be calculated using Eq.\eqref{eq: error_sig_v}, if needed. Note however, that it is fixed as before as detailed in Sec.~\ref{sec:calib_res_nonstock}. $\gamma$ and its standard error are the same as for the linear model as $\gamma$ is fitted and fixed ex ante.

\begin{table}[ht]
    \centering
    \rowcolors{2}{white}{gray!20}
    %\newcommand{\mystrut}{\rule{0pt}{2.25ex}} % Padding
    %\begin{tabular}{>{\mystrut}l|>{\centering\arraybackslash}p{1.3cm}|>{\centering\arraybackslash}p{1.3cm}|>{\centering\arraybackslash}p{1.3cm}|>{\centering\arraybackslash}p{1.3cm}|>{\centering\arraybackslash}p{1.3cm}|>{\centering\arraybackslash}p{1.3cm}|>{\centering\arraybackslash}p{1.3cm}|>{\centering\arraybackslash}p{1.3cm}}
    \begin{tabular}{l|c|c|c|c|c|c|c|c}
        
        {} & $\kappa$ & $\kappa_3$ & $\beta$ & $\gamma$ & $\sigma_N$ & $\sigma_V$ & $v_0$ & $\bar{\mathcal{L}}$ \\
        \hline
        \textbf{US}      & -0.002 & 0.222 & 0.099 & 4.17 & 0.042 & 0.011 & -0.001 & 1.73 \\
        \textbf{UK}      & -0.014 & 0.481 & 0.140 & 4.10 & 0.036 & 0.009 & 0.154 & 1.92 \\
        \textbf{AU}      & -0.055 & 1.30 & 0.134 & 4.24 & 0.037 & 0.010 & 0.231 & 1.85 \\
        \textbf{CH}      & -0.009 & 0.419 & 0.123 & 4.23 & 0.043 & 0.012 & 0.036 & 1.71 \\
        \textbf{JP}      & -0.029 & 0.318 & 0.133 & 4.15 & 0.076 & 0.015 & 0.075 & 1.41 \\
        \textbf{CA}      & 0.006 & 0.328 & 0.110 & 4.14 & 0.0443 & 0.012 & 0.038 & 1.68 \\
        \textbf{DE}      & -0.016 & 0.428 & 0.151 & 4.28 & 0.043 & 0.012 & 0.061 & 1.71  \\ \hline
        \textbf{SUGAR}   & -0.018 & 0.100 & 0.138 & 2.25 & 0.073 & 0.020 & -0.027 & 1.19 \\
        \textbf{CORN}    & -0.153 & 0.994 & 0.224 & 2.32 & 0.097 & 0.034 & 0.315 & 0.838 \\
        \textbf{LCATTLE} & 0.0277 & 1.87 & 0.193 & 2.20 & 0.059 & 0.013 & -0.055 & 1.39 \\
        \textbf{WHEAT}   & -0.045 & 1.07 & 0.200 & 2.52 & 0.087 & 0.025 & -0.097 & 0.975 \\
        \textbf{COPPER}  & -0.093 & 1.73 & 0.168 & 2.10 & 0.057 & 0.017 & 0.0315 & 1.39 \\
        \textbf{NATGAS}  & -0.093 & 2.09 & 0.252 & 2.32 & 0.142 & 0.048 & 0.112 & 0.396 \\
        \textbf{CRUDE}   & 0.087 & 0.602 & 0.276 & 2.77 & 0.107 & 0.030 & -0.368 & 0.812 \\ \hline
        \textbf{USBND}   & 0.004 & 0.505 & 0.088 & 5.32 & 0.046 & 0.003 & -0.000 & 1.65 \\
        \textbf{UKBND}   & 0.045 & 0.582 & 0.079 & 5.38 & 0.048 & 0.003 & 0.057 & 1.60 \\
        \textbf{CHBND}   & 0.012 & 0.211 & 0.025 & 6.06 & 0.086 & 0.006 & -0.070 & 1.56 \\
        \textbf{JPBND}   & 0.017 & 0.931 & 0.074 & 3.85 & 0.045 & 0.003 & 0.063 & 1.02 \\
        \textbf{AUBND}   & -0.024 & 0.764 & 0.074 & 4.74 & 0.039 & 0.003 & 0.097 & 1.65 \\
        \textbf{CABND}   & 0.004 & 1.44 & 0.120 & 4.57 & 0.044 & 0.003 & -0.034 & 1.90 \\
        \textbf{DEBND}   & -0.021 & 0.876 & 0.144 & 4.46 & 0.071 & 0.010 & 0.042 & 1.68 \\ \hline
        \textbf{CHFUSD}  & 0.007 & 0.682 & 0.042 & 5.46 & 0.043 & 0.005 & -0.092 & 1.93 \\
        \textbf{JPYUSD}  & 0.008 & 0.482 & 0.061 & 5.47 & 0.032 & 0.006 & -0.030 & 2.00 \\
        \textbf{AUDUSD}  & -0.034 & 1.89 & 0.048 & 7.01 & 0.049 & 0.005 & -0.056 & 1.99 \\
        \textbf{GBPUSD}  & 0.0165 & 0.545 & 0.055 & 7.07 & 0.044 & 0.005 & -0.001 & 2.10 \\
        \textbf{CADUSD}  & -0.009 & 0.616 & 0.102 & 7.76 & 0.019 & 0.003 & -0.025 & 2.51 \\
        \textbf{EURUSD}  & -0.022 & 0.601 & 0.037 & 6.40 & 0.032 & 0.005 & -0.091 & 2.00 \\
    \end{tabular}
    \caption{Same as Table~\ref{tab:4classes_Chiarella_lin} but for the de-drifted version of the non-linear model~\eqref{eq: ModifiedChiarellaNonlinear}.}
    \label{tab:4classes_Chiarella_nonlin}
\end{table}

\begin{table}[ht]
    \centering
    \rowcolors{2}{white}{gray!20}
    %\newcommand{\mystrut}{\rule{0pt}{2.25ex}} % Padding
    %\begin{tabular}{>{\mystrut}l|>{\centering\arraybackslash}p{1.3cm}|>{\centering\arraybackslash}p{1.3cm}|>{\centering\arraybackslash}p{1.3cm}|>{\centering\arraybackslash}p{1.3cm}|>{\centering\arraybackslash}p{1.3cm}}
    \begin{tabular}{l|c|c|c|c|c}
        %\toprule
        {} & $\Delta\kappa$ & $\Delta\kappa_3$ & $\Delta\beta$ & $\Delta\sigma_N$ & $\Delta v_0$ \\
        \hline
        \textbf{US}      & 0.001 & 0.081 & 0.021 & 0.001 & 0.001 \\
        \textbf{UK}      & 0.011 & 0.205 & 0.037 & 0.001 & 0.133 \\
        \textbf{AU}      & 0.023 & 0.325 & 0.028 & 0.001 & 0.071 \\
        \textbf{CH}      & 0.010 & 0.167 & 0.028 & 0.002 & 0.119 \\
        \textbf{JP}      & 0.012 & 0.110 & 0.032 & 0.002 & 0.122 \\
        \textbf{CA}      & 0.017 & 0.186 & 0.027 & 0.002 & 0.077 \\
        \textbf{DE}      & 0.009 & 0.120 & 0.022 & 0.001 & 0.093 \\ \hline
        \textbf{SUGAR}   & 0.009 & 0.049 & 0.044 & 0.002 & 0.027 \\
        \textbf{CORN}    & 0.019 & 0.149 & 0.073 & 0.010 & 0.061 \\
        \textbf{LCATTLE} & 0.027 & 0.793 & 0.068 & 0.002 & 0.074 \\
        \textbf{WHEAT}   & 0.028 & 0.729 & 0.066 & 0.008 & 0.088 \\
        \textbf{COPPER}  & 0.018 & 0.282 & 0.052 & 0.003 & 0.017 \\
        \textbf{NATGAS}  & 0.076 & 0.330 & 0.166 & 0.009 & 0.036 \\
        \textbf{CRUDE}   & 0.021 & 0.072 & 0.050 & 0.002 & 0.389 \\ \hline
        \textbf{USBND}   & 0.021 & 0.421 & 0.027 & 0.002 & 0.149 \\
        \textbf{UKBND}   & 0.027 & 0.320 & 0.038 & 0.002 & 0.001 \\
        \textbf{CHBND}   & 0.015 & 0.156 & 0.021 & 0.002 & 0.087 \\
        \textbf{JPBND}   & 0.017 & 0.071 & 0.038 & 0.007 & 0.083 \\
        \textbf{AUBND}   & 0.036 & 0.511 & 0.034 & 0.002 & 0.075 \\
        \textbf{CABND}   & 0.023 & 0.601 & 0.043 & 0.001 & 0.147 \\
        \textbf{DEBND}   & 0.030 & 0.790 & 0.038 & 0.002 & 0.025 \\ \hline
        \textbf{CHFUSD}  & 0.026 & 0.351 & 0.025 & 0.002 & -0.001 \\
        \textbf{JPYUSD}  & 0.011 & 0.373 & 0.023 & 0.004 & 0.072 \\
        \textbf{AUDUSD}  & 0.023 & 1.440 & 0.039 & 0.002 & 0.001 \\
        \textbf{GBPUSD}  & 0.030 & 0.323 & 0.024 & 0.001 & 0.111 \\
        \textbf{CADUSD}  & 0.025 & 0.457 & 0.082 & 0.001 & 0.061 \\
        \textbf{EURUSD}  & 0.021 & 0.362 & 0.024 & 0.001 & 0.100 
        %\bottomrule
    \end{tabular}
    \caption{Standard errors for the estimated parameters of Table~\ref{tab:4classes_Chiarella_nonlin}.}
    \label{tab:calib_errors_4classes_nonlin}
\end{table}

\begin{figure}[H]
    \centering
    \includegraphics[width=0.6\linewidth]{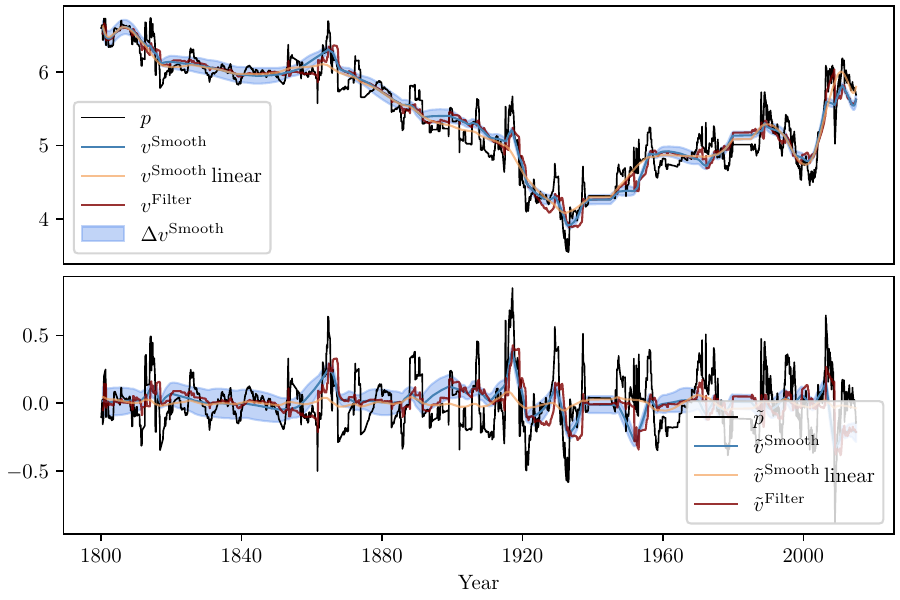}
    \caption{Same as Fig.~\ref{fig:US_calib_nonlin} but for the evolution of the price of the commodity copper.}
    \label{fig:Copper_calib_nonlin}
\end{figure}

\begin{figure}[H]
    \centering
    \includegraphics[width=0.6\linewidth]{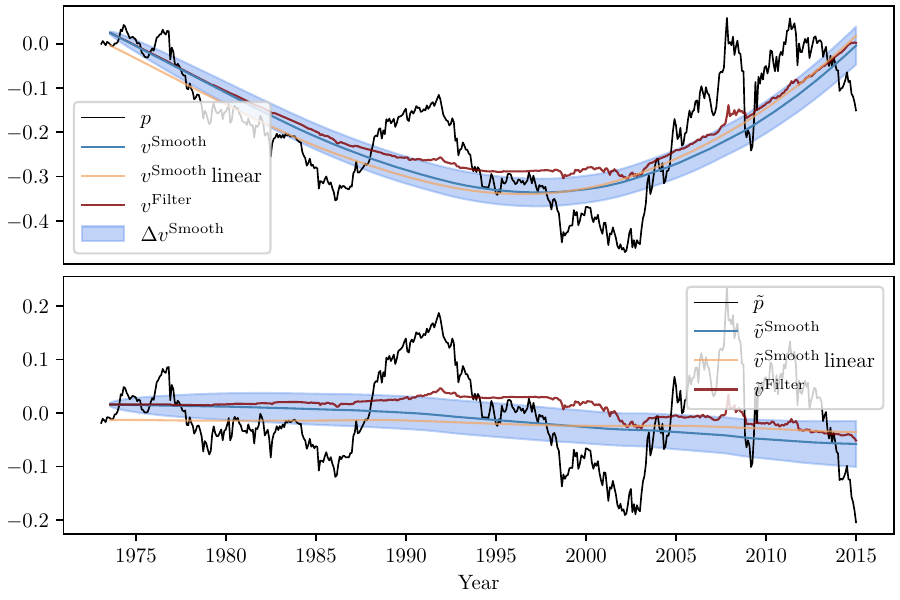}
    \caption{Same as Fig.~\ref{fig:US_calib_nonlin} but for the Canadian dollar vs. US dollar currency pair.}
    \label{fig:CADUSD_calib_nonlin}
\end{figure}

\begin{figure}[H]
    \centering
    \includegraphics[width=0.6\linewidth]{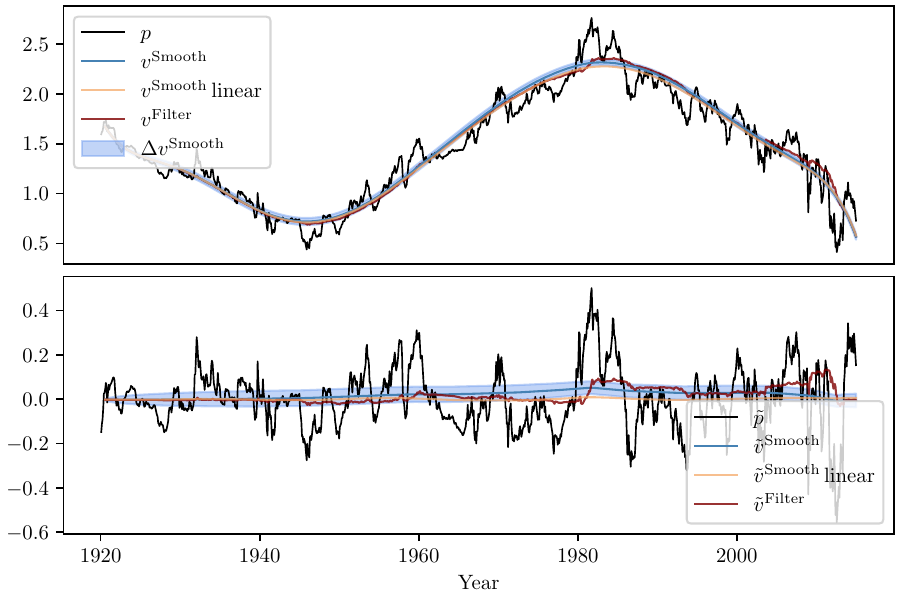}
    \caption{Same as Fig.~\ref{fig:US_calib_nonlin} but for the US government bond.}
    \label{fig:USBND_calib_nonlin}
\end{figure}

For the non-linear model the analogous results are visualised in Fig.~\ref{fig:US_calib_nonlin}, Fig.~\ref{fig:Copper_calib_nonlin}, Fig.~\ref{fig:CADUSD_calib_nonlin} and Fig.~\ref{fig:USBND_calib_nonlin}.


\subsection{Case Study: Insensitivity of Results to Changes in the Polynomial Order of the Drift}\label{sec: dedrift_change_robustness}
On the example of the US stock index, it is demonstrated that the results in this paper are not sensitive to the exact choice of the polynomial order used to infer the integrated drift $G_t$. This is important because the used heuristic was -- even if grounded on economic reason -- a modelling \textit{choice}. In Sec.~\ref{sec:calibration} the heuristic of an increase of polynomial order by one per decade of data was deployed. For the US index data this meant a polynomial of order 22. To show that the results subsist when changing this notion of drift, we repeat the calibration on a dedrifted log-price series that is dedrifted by polynomials of order $k=22 \pm 8 = \{14, \, 30\}$. For the dependence of $\sigma_V$ on $\sigma_N$, the $\Sigma$-factor listed in Tab.~\ref{tab:asset_factors} is recycled.
Tab.~\ref{tab:dedrifting_US_different_calib} shows that the results are not sensitive to the precise polynomial order as increasing the polynomial order from 22 to 30 or decreasing it to 14 barely changes the calibrated parameters, confirming the robustness of the proposed calibration scheme.
\begin{table}[ht]
    \centering
    %\rowcolors{2}{white}{gray!20}
    % Define a command to add padding
    %\newcommand{\mystrut}{\rule{0pt}{2.25ex}} % Adjust as needed for height
    % Set table width using p{width}
    %\begin{tabular}{>{\mystrut}l||>{\centering\arraybackslash}p{1.3cm}|>{\centering\arraybackslash}p{1.3cm}|>{\centering\arraybackslash}p{1.3cm}|>{\centering\arraybackslash}p{1.3cm}|>{\centering\arraybackslash}p{1.3cm}|>{\centering\arraybackslash}p{1.3cm}|>{\centering\arraybackslash}p{1.3cm}}
    \begin{tabular}{l||c|c|c|c|c|c|c}
        \toprule
        $k$ &  $\kappa$ &   $\beta$ &  $\gamma$ &  $\sigma_N$ &  $\sigma_V$ & $v_0$ & $\bar{\mathcal{L}}$ \\
        \midrule
        14      &  0.025 &  0.078 &  4.168 &    0.043 &    0.011 &  0.015 & 1.721 \\
        22      &  0.027 &  0.076 &  4.168 &    0.043 &    0.011 &  0.017 & 1.723 \\
        30      &  0.030 &  0.077 &  4.238 &    0.043 &    0.011 &  -0.009 & 1.724 \\
        
        \bottomrule
    \end{tabular}
    \caption{Calibration results for the fit on the de-drifted linear model~\eqref{eq:system_discretised_de-drifted} for the US stock index for where the dedrifting is performed with different polynomials of order $k$.}
    \label{tab:dedrifting_US_different_calib}
\end{table}

Tab.~\ref{tab:dedrifting_US_different_calib_NL} shows the same experiment but for the non-linear model~\eqref{eq: ModifiedChiarellaNonlinear}. For polynomial order $k=14$ and 22 the results are almost the same. For $k=30$, $\kappa$ is a bit smaller but the results are still robust.
\begin{table}[ht]
    \centering
    %\rowcolors{2}{white}{gray!20}
    %\newcommand{\mystrut}{\rule{0pt}{2.25ex}} % Padding
    %\begin{tabular}{>{\mystrut}l||>{\centering\arraybackslash}p{1.3cm}|>{\centering\arraybackslash}p{1.3cm}|>{\centering\arraybackslash}p{1.3cm}|>{\centering\arraybackslash}p{1.3cm}|>{\centering\arraybackslash}p{1.3cm}|>{\centering\arraybackslash}p{1.3cm}|>{\centering\arraybackslash}p{1.3cm}|>{\centering\arraybackslash}p{1.3cm}}
    \begin{tabular}{l||c|c|c|c|c|c|c|c}
        \toprule
        $k$ & $\kappa$ & $\kappa_3$ & $\beta$ & $\gamma$ & $\sigma_N$ & $\sigma_V$ & $v_0$ & $\bar{\mathcal{L}}$ \\
        \midrule
        14      & -0.004 & 0.202 & 0.099 & 4.17 & 0.043 & 0.011 & 0.000 & 1.73 \\
        22       & -0.002 & 0.222 & 0.099 & 4.17 & 0.042 & 0.011 & -0.001 & 1.73 \\
        30      & -0.017 & 0.213 & 0.103 & 4.17 & 0.042 & 0.011 & 0.016 & 1.73 \\
        \bottomrule
    \end{tabular}
    \caption{Same as Table~\ref{tab:dedrifting_US_different_calib} but for the non-linear model.}
    \label{tab:dedrifting_US_different_calib_NL}
\end{table}

Therefore, the results of this paper are overall not sensitive to the exact order $k$ of the polynomial with which the log-prices are de-drifted ex ante: results are robust.

\subsection{Sharpe Ratios from Model-implied Trend \& Value Signals} \label{app:Sharpes}
As supplementary information, we further provide the Sharpe ratios (SR) that could be obtained from the trend and value signals implied by model \eqref{eq: ModifiedChiarellaNonlinear}.
The trend and value signals are based on the de-drifted log-returns (recall that lower case letters stand for discretised quantities and tildes signify de-drifted quantities)
\begin{align}
    \tilde{s}^\textup{Trend}_t &= \beta \tanh (\gamma \tilde{m}_t),\\
    s^\textup{Value}_t &= \kappa (\tilde{v}_t - \tilde{p}_t) + \kappa_3 (\tilde{v}_t - \tilde{p}_t)^3,
\end{align}
where in this case $\tilde{v}$ is the \textit{filtered} fundamental log-value from the calibration of the non-linear model as investors can only know the filtered value (taking only past and present information into account) and not the smoothed (which also contains future information and thus can only be determined ex post).
Additionally, the trend signal is normalised with its own shifted and exponentially weighted moving standard deviation with a decay rate of $\alpha=1/7$, $\sigma^\textup{Trend}_{t-1}$, and clipped at $\pm 1$ std. dev. as is commonly done in practice:
\begin{equation}
    s^\textup{Trend}_t = \max\left(-1, \min\left(\frac{\tilde{s}^\textup{Trend}_t}{\sigma^\textup{Trend}_{t-1}}, 1\right)\right).
\end{equation}
Value signals are not normalised as fundamentalists take larger positions when mispricing is pronounced, instead of trading with constant risk. 

The profits and losses (PnL) of the signal based on the real (\textit{not} log-) prices $P_t$ are then given by
\begin{equation} \label{eq:PnLs}
    \text{PnL}_t^\textup{Trend/Value} = s_{t-1}^\textup{Trend/Value} \frac{P_t-P_{t-1}}{\sigma_{t-1}^P}
\end{equation}
where $\sigma_{t}^{P}$ is an exponentially weighted standard deviation of the price changes $P_t - P_{t-1}$ with a decay rate of $\alpha = 1/7$. Note that the trend signal has further been clipped at $\pm 1$ standard deviations as is commonly done in practice.
Using the definition of the SR, Eq.~\eqref{eq:Sharpe}, the the SRs of both strategies for two time periods of circa 60 years are calculated, listed in Tab.~\ref{tab:Sharpe_ratios}. Here, the SRs are calculated per asset class.%, that is, the PnLs within one asset class have been averaged.
\begin{table}[ht]
    \centering
    \begin{minipage}{0.4\linewidth}
    \centering
    %\begin{tabular}{
       % >{\mystrut}l|>{\centering\arraybackslash}p{2cm}|
       % >{\centering\arraybackslash}p{2cm}
    %}
    \begin{tabular}{l||c|c}
        \toprule
        Asset Class & SR Trend & SR Value \\
        \midrule
        % the commented out are the SR from non-normalised trend signal
        %\textbf{IDX}  & 0.741 & 0.201  \\
        %\textbf{CMD}  & 0.134 & 0.043  \\
        %\textbf{BND}  & 0.321 & 0.657  \\
        %\textbf{FXR}  & -     & -      \\
        \textbf{IDX}  & 0.946 & 0.201  \\
        \textbf{CMD}  & 0.091 & 0.043  \\
        \textbf{BND}  & 0.393 & 0.657  \\
        \textbf{FXR}  & -     & -      \\
        \bottomrule
    \end{tabular}
    \end{minipage}\hspace{0.02\linewidth}%\hfill
    \begin{minipage}{0.4\linewidth}
    \centering
    %\begin{tabular}{
     %   >{\mystrut}l|>{\centering\arraybackslash}p{2cm}|
      %  >{\centering\arraybackslash}p{2cm}
    %}
    \begin{tabular}{l||c|c}
        \toprule
        Asset Class & SR Trend & SR Value \\
        \midrule
        %\textbf{IDX}  & 0.501 & 0.165  \\
        %\textbf{CMD}  & -0.046 & 0.315  \\
        %\textbf{BND}  & 0.231 & 0.589  \\
        %\textbf{FXR}  & 0.208 & 0.349  \\
        \textbf{IDX}  & 0.659 & 0.165  \\
        \textbf{CMD}  & 0.053 & 0.315  \\
        \textbf{BND}  & 0.581 & 0.589  \\
        \textbf{FXR}  & 0.423 & 0.349  \\
        \bottomrule
    \end{tabular}
    \end{minipage}
    
    \caption{Sharpe ratios (SR) for the trend and value PnLs defined via Eq.~\eqref{eq:PnLs} and Eq.~\eqref{eq:Sharpe} and averaged within each of the four asset classes. Left: SR calculated on data from 1890--1950, right: data from 1950--end.}
    \label{tab:Sharpe_ratios}
\end{table}

The Sharpe ratios in Tab.~\ref{tab:Sharpe_ratios} yield reasonably good results considering that we did not really optimise for high SRs. Also note that the trend signal may be be weaker than other reported trend signals in the literature as the trend signal here is defined on the \textit{de-drifted} log-returns, or: the log-mispricing, taking out the long-only bias -- to that our model is not only de-drifted with a linear drift (as usually done in the literature, if done at all) but with a time-varying polynomial drift. As these drifts capture overall market growth and influx of investors into the market, which are not really what we understand as trending, it is only natural to have a lower Sharpe as this Sharpe is the expected gain \textit{in excess} of the long-only strategy. This de-drifting procedure is chosen to improve the estimation of fundamental value, one of the primary goals and outcomes of this study. Improving the value estimation and maximising the SR from trend signals poses a trade-off.

\section{Further Sloppiness Analysis Results}\label{app:sloppy}
Here we present and discuss further results of the sloppiness analysis for the four asset classes.

\begin{figure}[ht]
    \centering
    \includegraphics[width=0.9\linewidth]{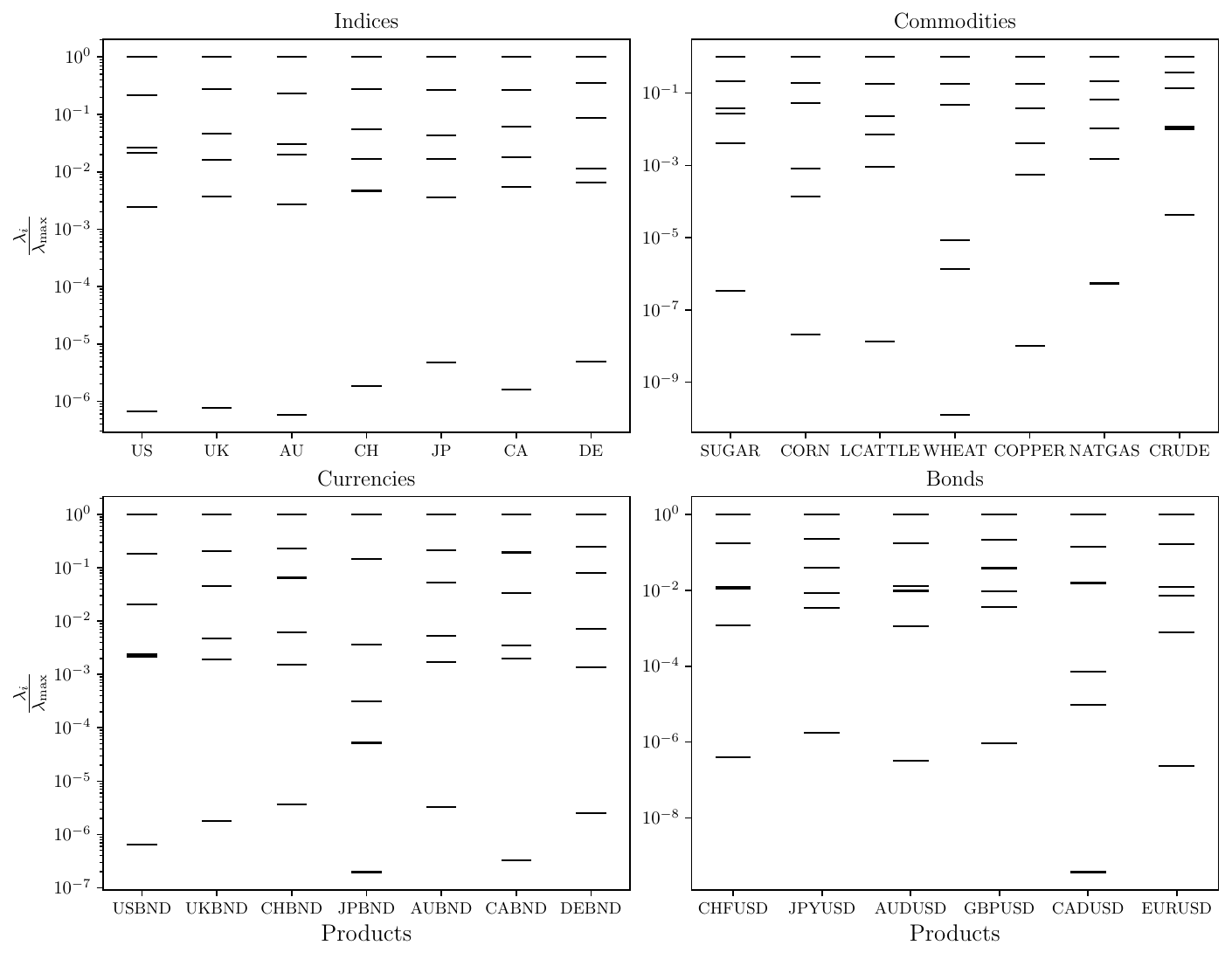}
    \caption{Sensitivity eigenvalue spectra due to the Hessian in eq.~\ref{eq:Hessian_sloppiness} per asset for the linear model. Each spectrum is normalised by its maximum eigenvalue for scale.}
    \label{fig:sloppy_linear_eigvals_spot_all}
\end{figure}

\begin{figure}[htbp]
    \centering
    \includegraphics[width=0.55\linewidth]{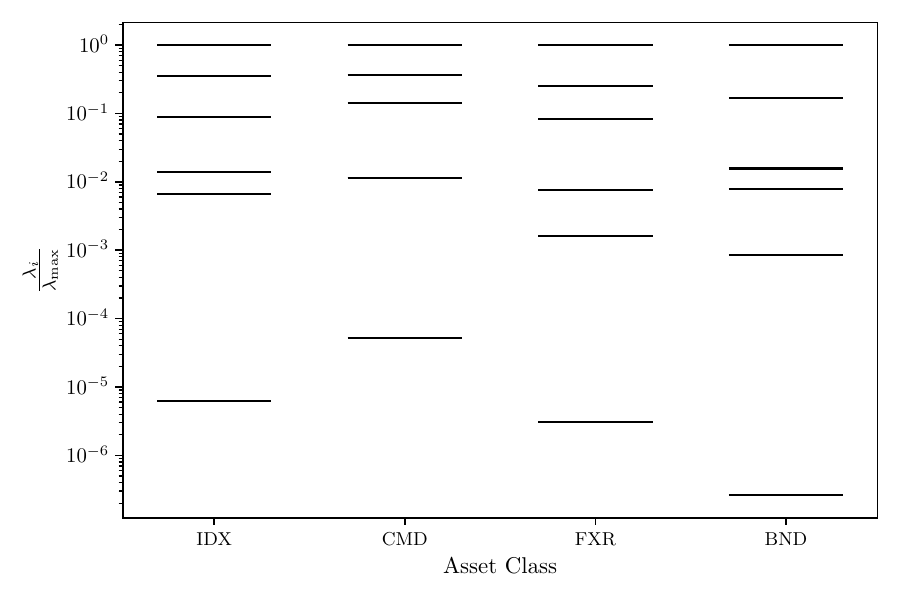}
    \caption{Same as Fig.~\ref{fig:sloppy_linear_eigvals_spot_all} but for a Hessian that is averaged per asset class (linear model).}
    \label{fig:sloppy_eigenvals_linear_byclass}
\end{figure}

\begin{figure}[htbp]
    \centering
    \includegraphics[width=0.49\linewidth]{images/sloppy_lin_mispr_eigvecs_IDX.pdf}
    \includegraphics[width=0.49\linewidth]{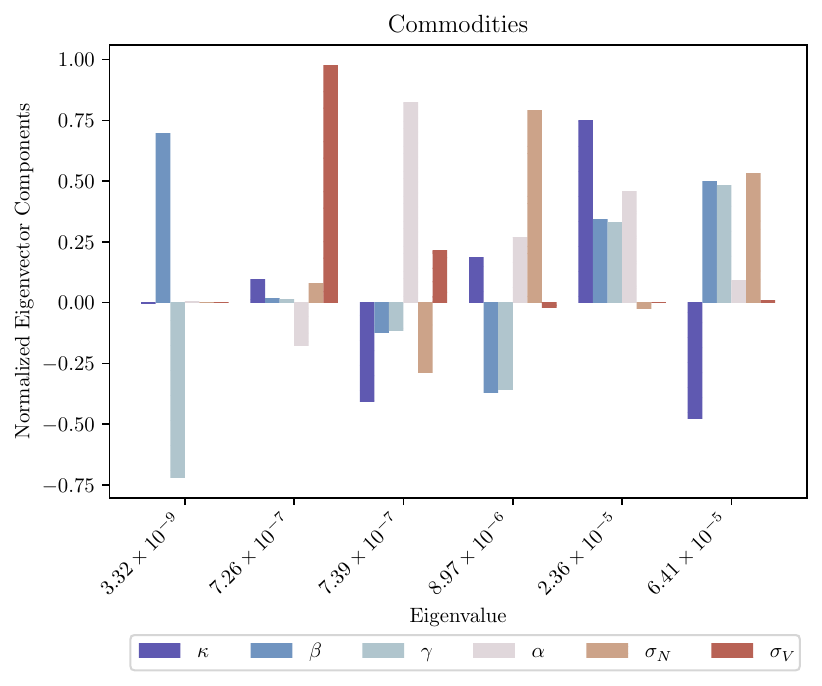} \\
    \includegraphics[width=0.49\linewidth]{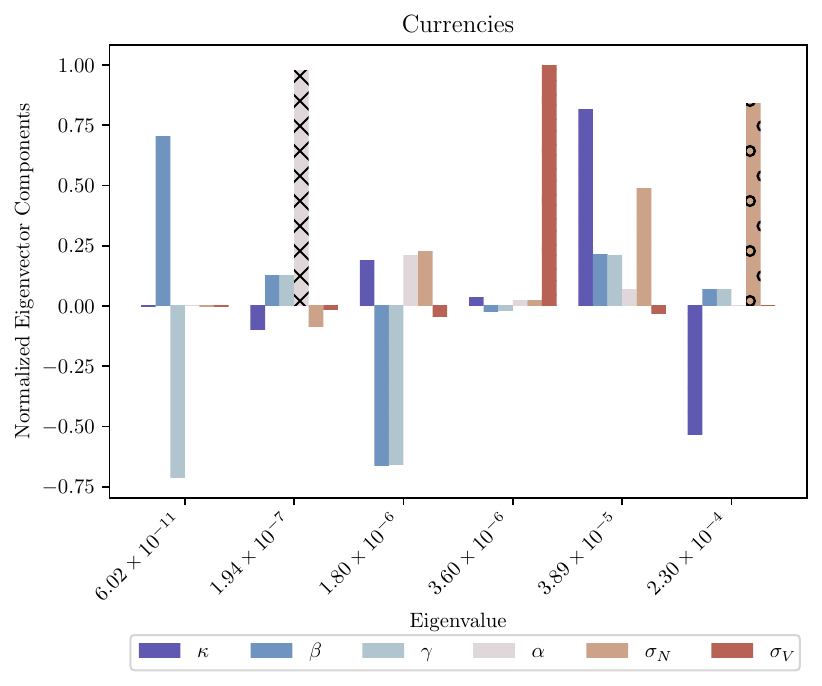}
    \includegraphics[width=0.49\linewidth]{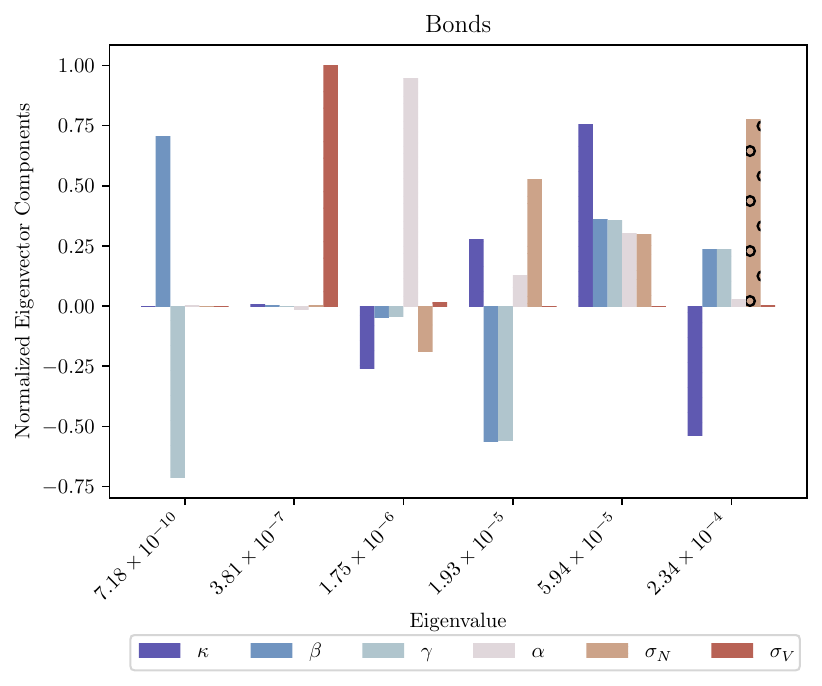}
    \caption{Normalised eigenvectors from the sloppy analysis (Sec.~\ref{sec:sloppy}) of the linear model on the mispricing $\delta=p-v$ for the four asset classes, ranked by eigenvalue magnitude. Within one asset class the Hessians of the individual assets have been averaged before the eigendecomposition was performed to get an in-class view.}
    \label{fig:sloppy_eigvecs_lin_byclass}
\end{figure}

The complete set of sensitivity spectra of the mispricing $\delta=p-v$ of the linear model~\eqref{eq: ModifiedChiarella} based on the eigenvalues of $\mathcal{H}$ in Eq.~\eqref{eq:Hessian_sloppiness}, depicted in groups of asset classes, are provided in Fig.~\ref{fig:sloppy_linear_eigvals_spot_all}. It is clearly visible that each spectrum spans many decades, providing strong evidence for the dynamical system describing the price and value processes being sloppy over all assets and asset classes. 
For completeness, we depict the sensitivity spectrum of the averaged Hessians $\mathcal{H}$, where the average is taken per asset class, in Fig.~\ref{fig:sloppy_eigenvals_linear_byclass}. These are the eigenvalues corresponding to the eigenvectors in Fig.~\ref{fig:sloppy_eigvecs_lin_byclass}
, which depicts the parameter eigendirections as discussed in Sec.~\ref{sec:sloppy} for the four averaged asset classes (indices, commodities, currencies, bonds).
Fig.~\ref{fig:sloppy_eigvecs_lin_byclass} shows that the qualitative results for the parameter eigendirections are quite universal over the the four asset classes. Their key characterisations as well as an interpretation of each mode are explained on the example of indices in Sec.~\ref{sec:sloppy}.

Comparing the first mode, the \textit{variance mode}, over all four asset classes from Fig.~\ref{fig:sloppy_eigvecs_lin_byclass}, it can be seen that they all have the same composition. The main difference lies in the magnitude of trend followers' (TF) contribution to the mispricing variance, which is the highest for commodities, where changes in $\beta\gamma$, which are the parameters associated with trend following, have the same power to deteriorate the fit than a blunt increase of noise traders through $\sigma_N$. Almost the same holds true for indices, where TFs have only slightly less power. TFs' impact on bond mispricings is attenuated in comparison, while it is evident that TF plays almost no role in currency markets.

The second, third and fifth mode in the eigendecomposition for indices, which are explained in Sec.~\ref{sec:sloppy}, occur as the second, third and fourth mode for asset classes commodities and bonds. This is because the fourth mode for the indices is the \textit{value mode}, indicating that value noise plays a larger role in index mispricings than in those of bonds and commodities. The largest relative impact has value on currencies, which may be explained through the extenuated TF effect. Otherwise, the currency modes are also similar to the other asset classes' modes.

The least significant mode, the \textit{trend saturation mode}, is again common to all four asset classes. In absolute terms it is again (as with the other TF contributions) most significant in commodities markets and least significant for currencies as can be read off from the respective eigenvalues belonging to mode six.

Analogously obtained sloppiness analysis results are shown for the non-linear model~\eqref{eq: ModifiedChiarellaNonlinear}.

\begin{figure}
    \centering
    \includegraphics[width=0.99\linewidth]{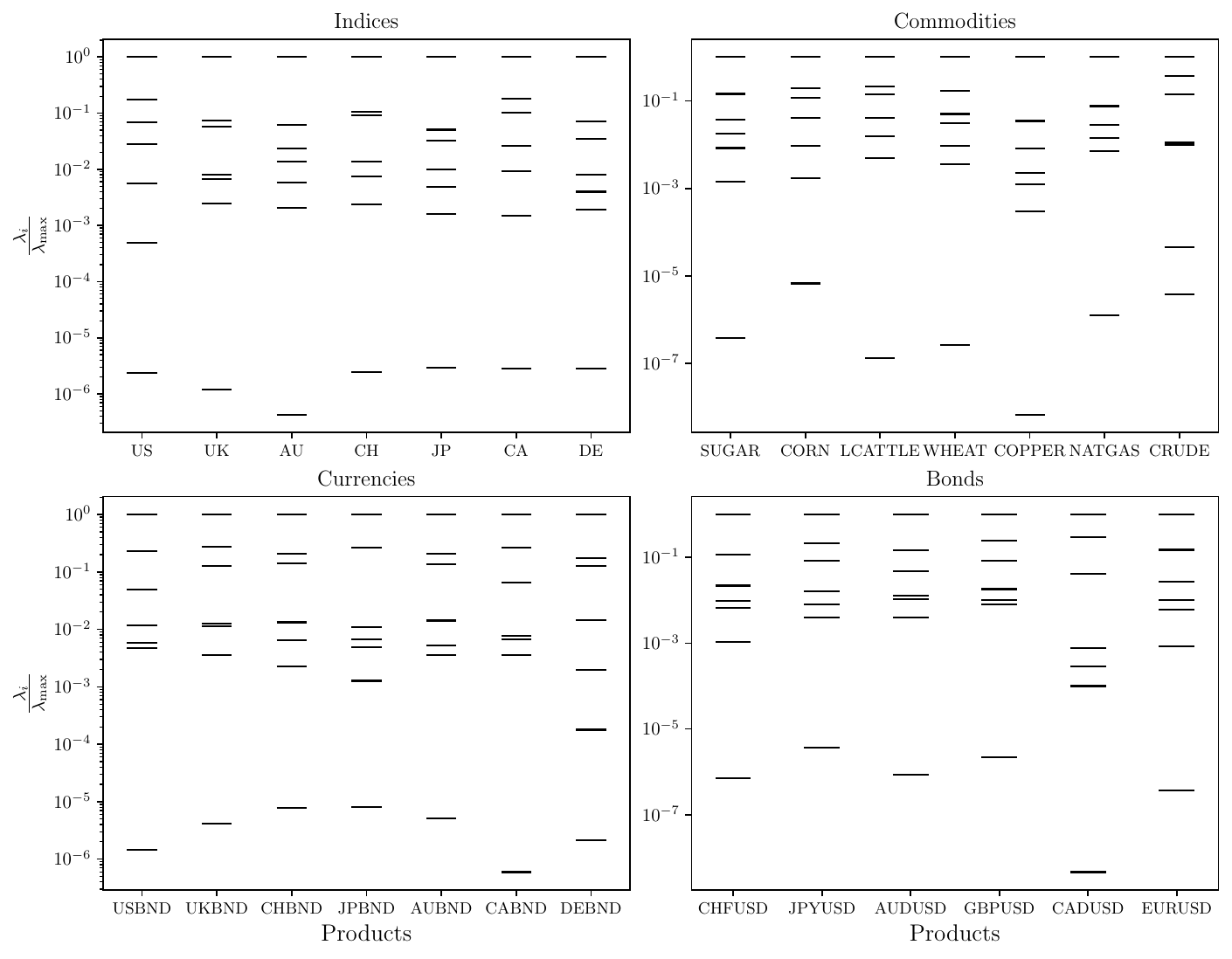}
    \caption{Same as Fig.~\ref{fig:sloppy_linear_eigvals_spot_all} but for the non-linear model and optimal parameters $\theta$ from Table~\ref{tab:4classes_Chiarella_nonlin}.}
    \label{fig:sloppy_NONlinear_eigvals_spot_all}
\end{figure}

Fig.~\ref{fig:sloppy_NONlinear_eigvals_spot_all} shows the sensitivity eigenvalue spectra for each product of the four asset classes grouped by asset class. As in Fig.~\ref{fig:sloppy_linear_eigvals_spot_all} for the linear model, the spectra span multiple decades, providing evidence that this model, too, is sloppy. Interestingly though, the non-linear model seems to be less sloppy than the linear one, even though one parameter has been added. This can be interpreted as the model output reacting less diversely in magnitude to small changes in different parameters, suggesting that this model is more stable in a sense.

\begin{figure}
    \centering
    \includegraphics[width=0.49\linewidth]{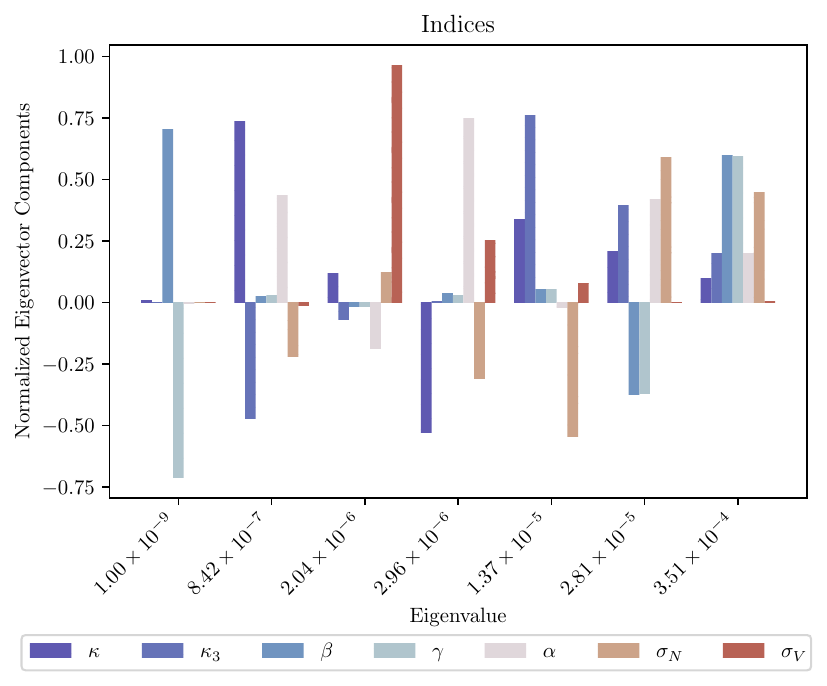}
    \includegraphics[width=0.49\linewidth]{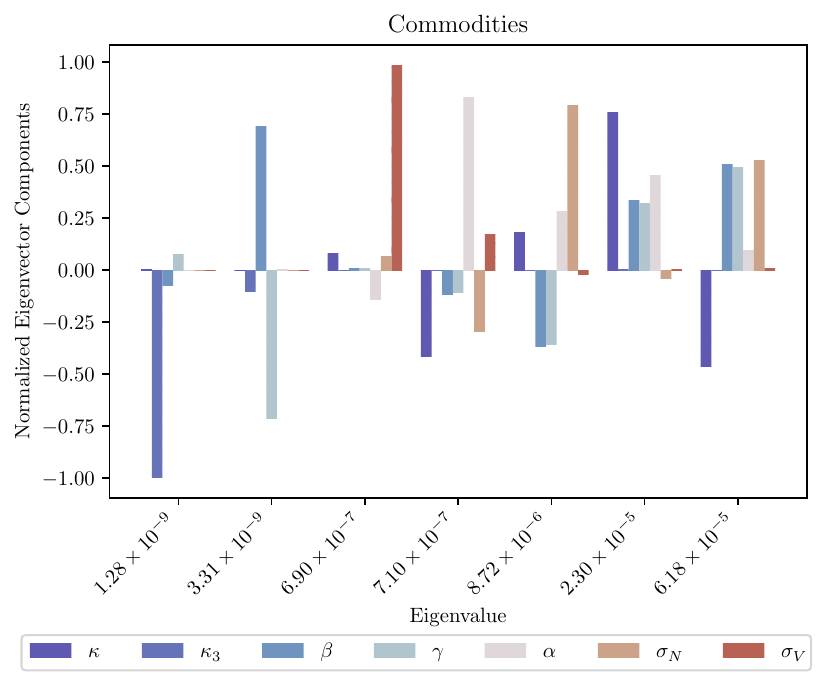} \\
    \includegraphics[width=0.49\linewidth]{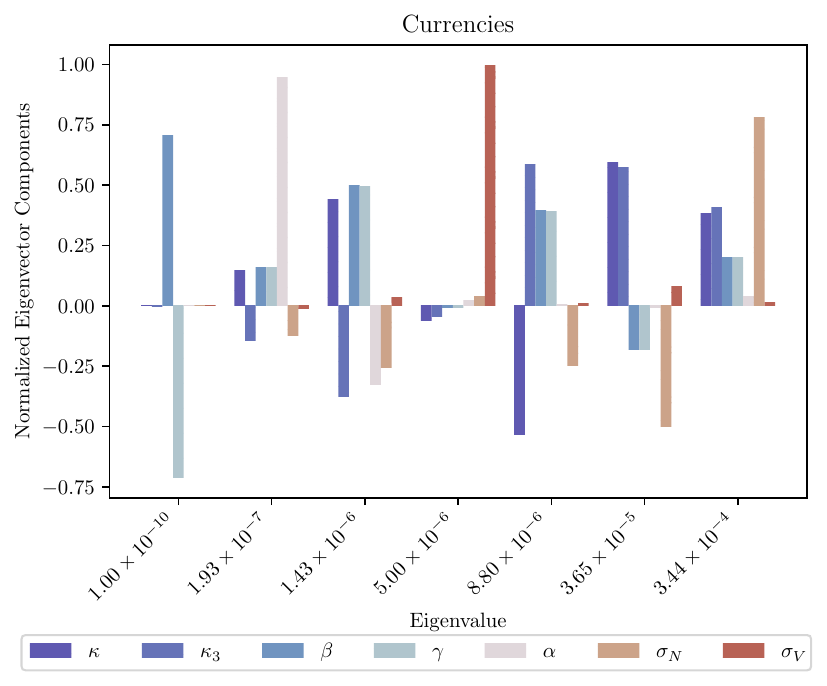}
    \includegraphics[width=0.49\linewidth]{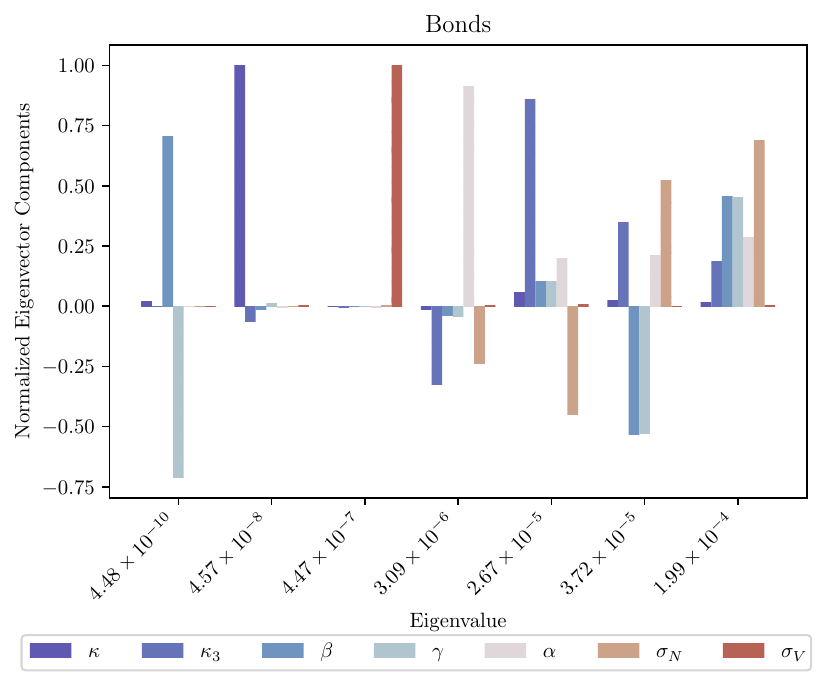}
    
    \caption{Same as Fig.~\ref{fig:sloppy_eigvecs_lin_byclass} but for the non-linear model and optimal parameters $\theta$ from Table~\ref{tab:4classes_Chiarella_nonlin}.}
    \label{fig:sloppy_eigvecs_NONlin_byclass}
\end{figure}

The same eigenvalue spectra for the Hessians that are averaged over one asset class are depicted in Fig.~\ref{fig:sloppy_eigvecs_NONlin_byclass}. Those are the normalised (by the spectral radius) eigenvalues belonging to the full eigendecomposition provided through Fig.~\ref{fig:sloppy_eigvecs_NONlin_byclass}, which, naturally, span multiple decades, too.
\begin{figure}
    \centering
    \includegraphics[width=0.6\linewidth]{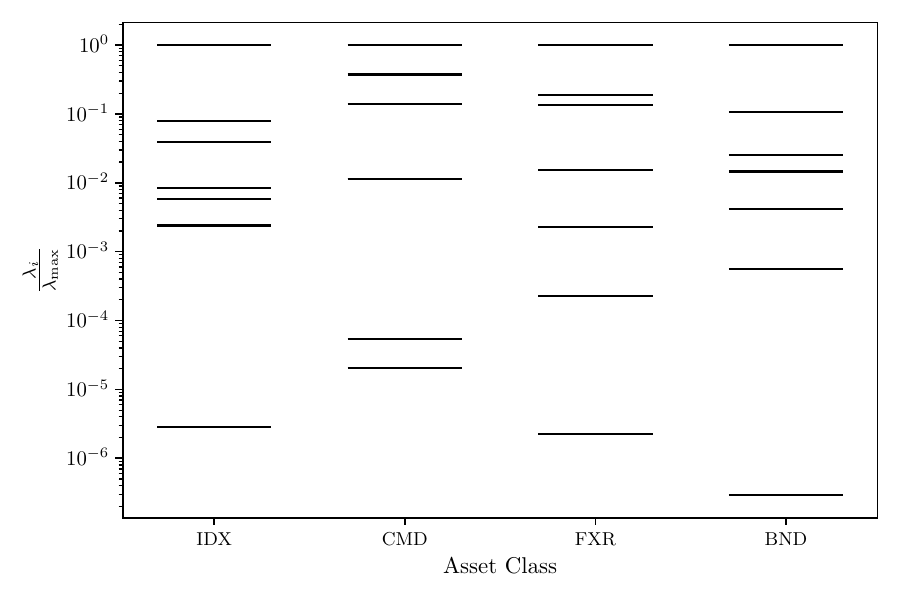}
    \caption{Same as Fig.~\ref{fig:sloppy_eigenvals_linear_byclass} but for the non-linear model and optimal parameters $\theta$ from Table~\ref{tab:4classes_Chiarella_nonlin}.}
    
\label{fig:sloppy_eigenvals_NONlinear_byclass}
\end{figure}

The statements made for the sloppiness analysis of the linear model in sec.~\ref{sec:sloppy} and before in this section in the context of Fig.~\ref{fig:sloppy_eigvecs_lin_byclass} are also true for the non-linear version as can be seen by comparing Fig.~\ref{fig:sloppy_eigvecs_lin_byclass} with Fig.~\ref{fig:sloppy_eigvecs_NONlin_byclass}. Remarkably, even the eigenvalues are of similar size for all asset classes and eigendirections, indicating that responses to small changes in certain parameters result in similar changes in outputs, which can be measured through the loss function, Eq.~\eqref{eq:loss_sloppiness}, and hence constitute similar degradations in fit. Small deviations are expected as the parameter space has changed between the two models and, indeed, has one more dimension in the non-linear case.

The first modes in Fig.~\ref{fig:sloppy_eigvecs_NONlin_byclass} can be identified as the second modes in sloppy analysis of the linear model (comp. Fig.~\ref{fig:sloppy_eigvecs_lin_byclass}), except for commodities where the order has not changed. While both modes either way control the dispersion of price and value, this change indicates that the model has become more sensitive to the magnitude of TFs. This may be explained by the non-linear model being closer to criticality (and also more sensitive with regards to the bifurcation), which is supported by the numerical studies of the modality of the non-linear system, which often shows weak levels of bimodality in the mispricing, which has been shown to correspond to the oscillatory dynamical phase, which was not present in the simulation study of the linear model using the optimal calibrated parameters, indicating that in the linear model the asset dynamics are farther away from reaching the oscillatory phase through a destabilising increase in TF. For commodities this is not the case, likely due to TFs being so strong that they are major constituents of most modes anyways.

A further difference between the non-linear and the linear eigendirections is that the additional direction in the non-linear case seems to be identifiable as a \textit{value mode}, which was not present in the linear counterparts (comp. Fig.~\ref{fig:sloppy_eigvecs_lin_byclass}). This value mode is sometimes composed of both contributions to the fundamentalists' demand, $\kappa$ and $\kappa_3$ (indices), in other cases, however, only by one of them (commodities and bonds); merely for currencies can this mode not readily be made out. On the other hand, close inspection of Table~\ref{tab:4classes_Chiarella_nonlin} reveals that in the case of commodities (where the eigendirection corresponds to $\kappa_3$) $\kappa$ was almost always negative, while for bonds (where the eigendirection corresponds to $\kappa$) $\kappa$ is always positive, which means that they potentially describe the exact same component, which in the case of bonds could not be further distinguished by the calibration method as both $\kappa$ and $\kappa_3$ are positive. This view is supported by the ill-constrainedness suggested by the low rank of the eigenvalue (and generally its magnitude) corresponding to this parameter direction.

\end{document}